\documentclass[manuscript]{acmart}

\usepackage{tabularx,makecell}

\usepackage{multirow}
\usepackage{cite}
\usepackage{booktabs}
\usepackage{wrapfig}
\usepackage{multirow}
\usepackage{colortbl}
\usepackage{braket}
\usepackage{tcolorbox}

\usepackage{amsmath,amssymb,amsfonts}
\usepackage{mathtools}
\usepackage{algorithm}
\usepackage{algpseudocode}
\usepackage{comment}
\usepackage{tabularray}
\usepackage{graphicx}
\usepackage{textcomp}
\usepackage{xcolor}
\usepackage{textcomp}
\usepackage{xcolor}
\usepackage{hyperref}
\usepackage{graphics}
\usepackage{caption}
\usepackage{pifont}
\usepackage{fvextra}
\usepackage{subcaption}
\usepackage{flushend}
\RequirePackage{color}
\definecolor{RED}{rgb}{1,0,0}
\definecolor{BLUE}{rgb}{0,0,1}
\definecolor{White}{rgb}{1,1,1}

\newcommand{\ie}{\textit{i}.\textit{e}.}
\newcommand{\eg}{\textit{e}.\textit{g}.}

\usepackage{adjustbox}

\AtBeginDocument{%
  }

\begin{document}

\newcommand{\our}{Q-Bridge~}

\title{Q-Bridge: Code Translation for Quantum Machine Learning via LLMs}

\author{Runjia Zeng\textsuperscript{\dag}}

\email{rz4545@rit.edu}

\affiliation{%
\institution{Rochester Institute of Technology}
\city{Rochester}
\country{USA}
}

\author{Priyabrata Senapati\textsuperscript{\dag}}

\email{psenapat@kent.edu}

\affiliation{%
\institution{Kent State University}
\city{Kent}
\country{USA}
}

\thanks{\textsuperscript{\dag}Equal contribution.}

\author{Ruixiang Tang}
\affiliation{%
  \institution{Rutgers University}
  \city{University–New Brunswick}
  \country{USA}}
  \email{ruixiang.tang@rutgers.edu}

\author{Dongfang Liu\textsuperscript{*}}  
\affiliation{%
\institution{Rochester Institute of Technology}  
\city{Rochester}  
\country{USA}}  
\email{dongfang.liu@rit.edu}  

\author{Qiang Guan\textsuperscript{*}}  
\affiliation{%
\institution{Kent State University}  
\city{Kent}  
\country{USA}}  
\email{qguan@kent.edu}  

\thanks{\textsuperscript{*}Corresponding authors.}





\renewcommand{\shortauthors}{Zeng et al.}

\begin{abstract}
Large language models have recently shown potential in bridging the gap between classical machine learning and quantum machine learning. However, the lack of standardized, high-quality datasets and robust translation frameworks limits progress in this domain. We introduce Q-Bridge, an LLM-guided code translation framework that systematically converts CML implementations into executable QML variants. Our approach builds on a self-involving pipeline that iteratively expands a verified seed codebase into a large-scale dataset, CML-2-QML, integrating verifiable and unverifiable code pairs. The Q-Bridge model is fine-tuned using supervised LoRA adaptation for scalable and memory-efficient training, achieving faithful and interpretable quantum code generation across diverse architectures. Empirical analysis confirms the feasibility of direct CML-to-QML translation and reveals consistent structural alignment between classical and quantum paradigms. Case studies further demonstrate that Q-Bridge can maintain deterministic correctness and also enable creative architectural exploration. This work establishes the first reproducible framework and dataset for LLM-driven quantum code translation, offering a foundation for scalable quantum AI development.
\end{abstract}

\ccsdesc[500]{Software and its engineering~Application specific development environments}
\keywords{Large Language Models, Quantum Computing, Variational Quantum Algorithm, Quantum Machine Learning, Code Generation, Quantum Circuit Generation, Classical Machine Learning}


\maketitle
\includegraphics[height=1.3em]{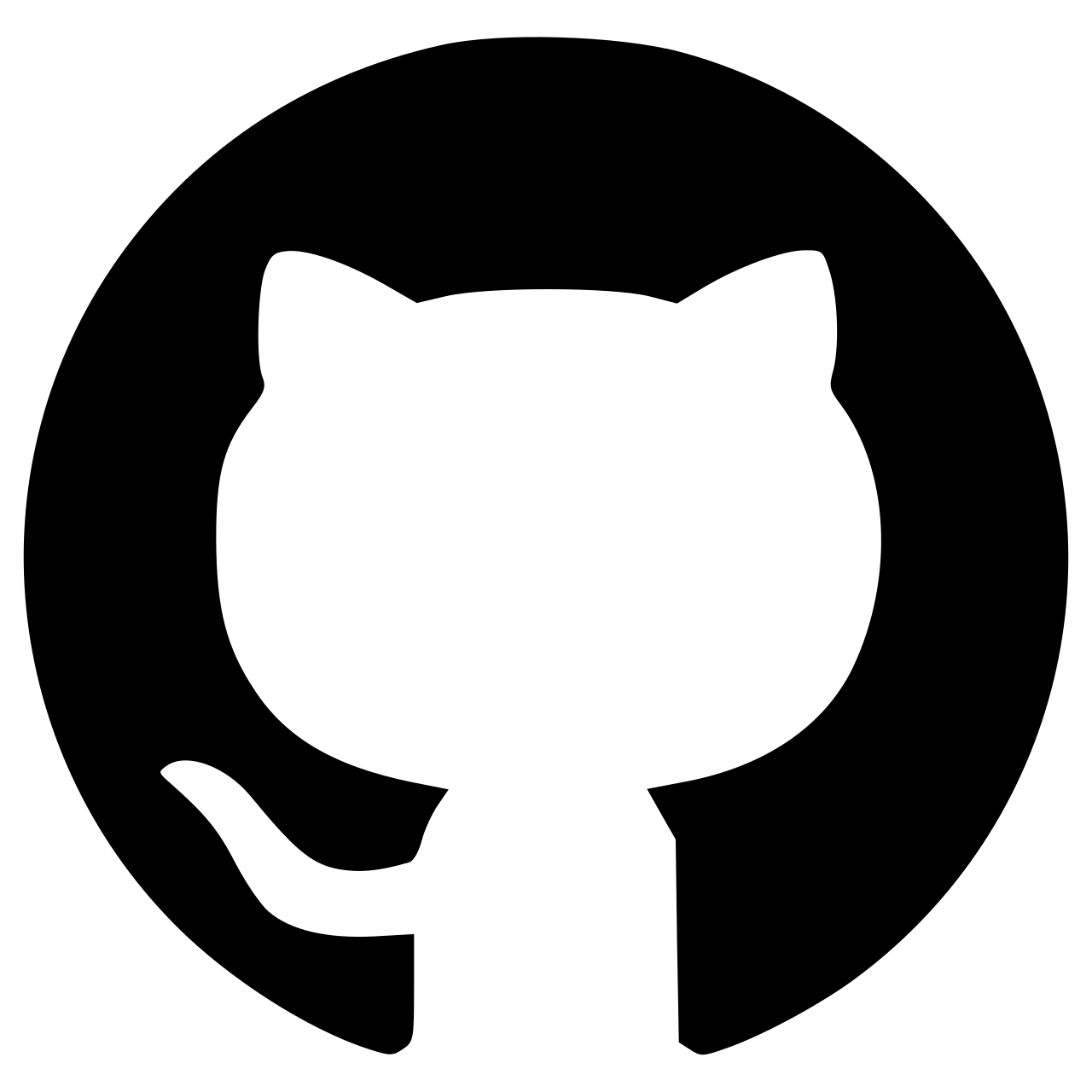} Code: \href{https://github.com/runtsang/Q-Bridge}{https://github.com/runtsang/Q-Bridge}\\
\includegraphics[height=1.3 em]{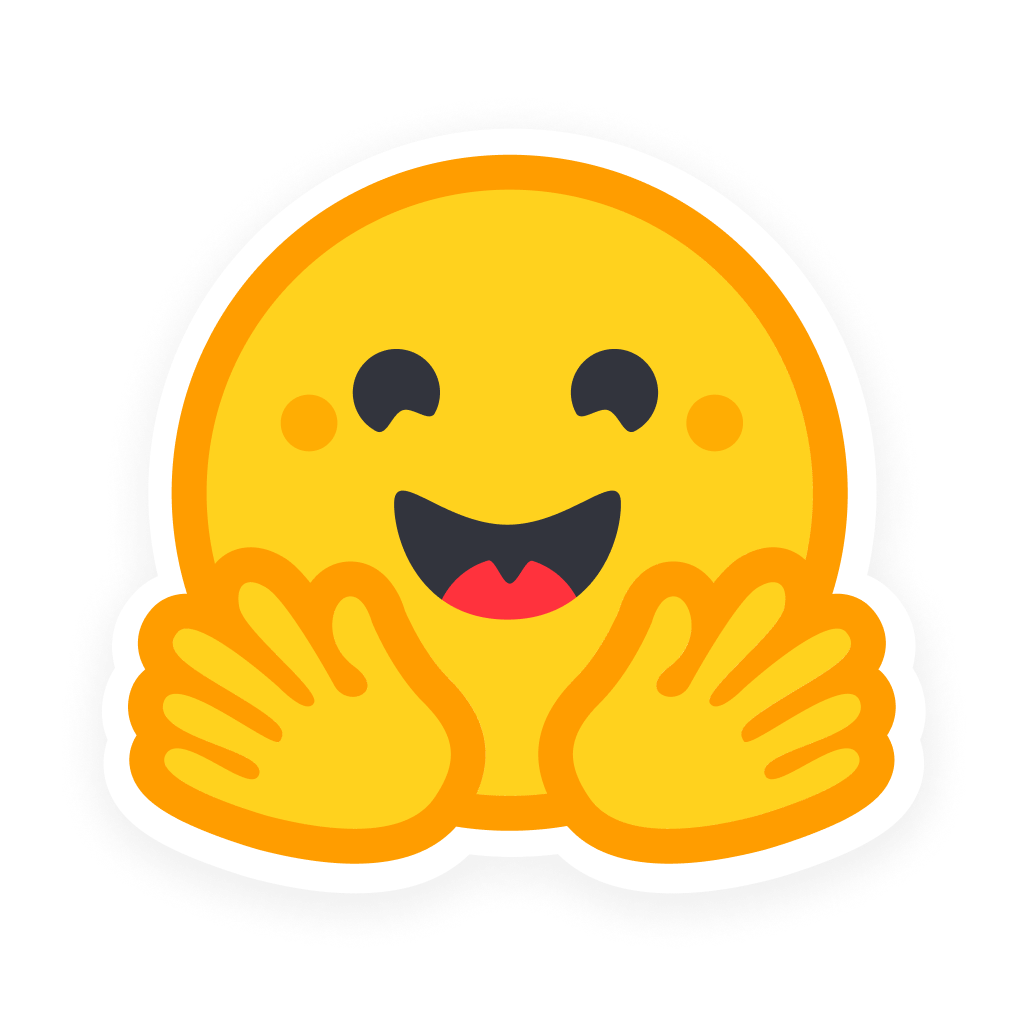} CML-2-QML Dataset: \href{https://huggingface.co/datasets/runjiazeng/CML-2-QML}{https://huggingface.co/datasets/runjiazeng/CML-2-QML}\\
\includegraphics[height=1.3em]{figures/hf-logo.png} Q-Bridge Model: \href{https://huggingface.co/runjiazeng/Q-Bridge}{https://huggingface.co/runjiazeng/Q-Bridge}
\section{Introduction}

Quantum computing offers a fundamentally different computational model based on superposition, entanglement, and interference, enabling algorithmic behaviors that are not efficiently available to classical hardware. Canonical examples include Shor's algorithm for integer factorization with exponential speedups and Grover's algorithm for unstructured search with quadratic speedups \citep{shor1994algorithms,grover1996fast}. Experimental milestones such as large scale random circuit sampling on superconducting platforms have shown devices executing sampling tasks believed to be classically prohibitive \citep{arute2019quantum}. Despite this promise, contemporary devices remain in the noisy intermediate scale quantum (NISQ) regime, with limited qubit counts, short coherence times, restricted connectivity, and imperfect control. Programming such hardware requires circuit depth minimization, device aware compilation, and noise robust evaluation strategies, \ie, as a result, practical algorithm design demands expertise in both quantum information and systems engineering. At the same time, industrial roadmaps project steadily scaling qubit counts and improved control stacks, motivating methods that extract learning value from near term hardware while remaining forward compatible with future systems \citep{preskill2018quantum,ibmfutureroadmap}.

\begin{figure*}
\vspace{-2mm}
\centering
\includegraphics[width=1.0\textwidth]
    {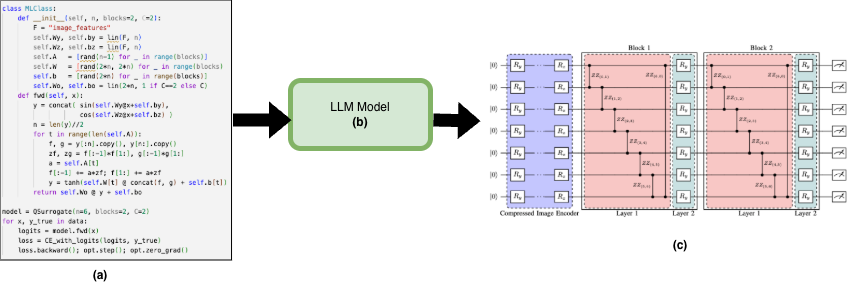}
  \caption{Classical to quantum translation pipeline that includes (a) conventional ML code is provided as an input, (b) a LLM interprets the classical ML code and synthesizes an equivalent quantum ML specification, (c) the generated design is instantiated as a parameterized quantum circuit encoder plus variational blocks with single qubit rotations, ZZ entanglers, and measurements ansatz \citep{senapati2024pqml}.}
  \Description{Diagram showing the translation pipeline from classical ML code to quantum ML specification using LLMs, resulting in a quantum circuit design.}
\label{fig:qnn_blocks}
\vspace{-5mm}
\end{figure*}

Variational quantum algorithms (VQAs) have emerged as a leading strategy to harness NISQ devices \citep{cerezo2021variational}. VQAs couple a parameterized quantum circuit (PQC) with a classical optimizer to minimize a cost function measured on hardware. This hybrid loop delegates gradient based or gradient free search to classical resources while reserving state preparation and measurement to the quantum device. Foundational instances include the Variational Quantum Eigensolver (VQE) for quantum chemistry and materials and the Quantum Approximate Optimization Algorithm (QAOA) for combinatorial optimization \citep{peruzzo2014variational,farhi2022quantum}. VQAs are attractive because they can be tailored to device constraints (\eg, shallow depth, topology aware entanglers), and incrementally improved with error aware training heuristics. Nonetheless, open challenges persist, \ie, non-convex optimization under hardware noise, measurement overhead for accurate gradients, and trainability issues such as barren plateaus where regions of exponentially vanishing gradients that impede learning have been documented and partially characterized \citep{mcclean2018barren,cerezo2021variational}. These realities underline the need for principled design automation and robust workflow tooling when deploying VQAs for real tasks. For readers with a deep learning background, it is helpful to organize QML into three complementary paradigms: (i) quantum enhanced classical ML, in which quantum devices estimate inner products or linear algebra primitives for kernel and matrix subroutines \citep{havlivcek2019supervised,schuld2021supervised}, (ii) native quantum learning with parameterized models (variational classifiers (or regressors), generative models) \citep{perez2020data,cong2019quantum}, and (iii) quantum optimization and sampling primitives (\eg, QAOA variants) integrated into ML pipelines \citep{farhi2022quantum}. While demonstrations are modest in scale, these formulations probe whether quantum resources can alleviate classical bottlenecks such as expensive kernel evaluation, representation efficiency, or sampling variance \citep{biamonte2017quantum, preskill2018quantum,cerezo2021variational}. Concurrently, large language models (LLMs) have achieved strong performance on code understanding and generation benchmarks. Domain tuned efforts now target quantum stacks, \ie, the Qiskit code assistant, built on a granite family model, demonstrates substantial gains on quantum specific code generation tasks, including the Qiskit HumanEval benchmark \citep{dupuis2024qiskit}. Complementary studies evaluate LLMs as explanatory agents for quantum algorithms, suggesting a role in onboarding non-experts. The steep learning curve of quantum programming reversible logic, measurement semantics, hardware aware transpilation forms a barrier for the classical ML community. We posit that LLMs can serve as bridges between classical ML (CML) workflows and quantum ML (QML) implementations by (i) mapping classical data pipelines to quantum feature encoders or measurement layouts, (ii) proposing task appropriate QML paradigms (kernel vs. PQC vs. sampling), (iii) generating device-aware circuits and training loops in Qiskit or PennyLane, and (iv) explaining the design choices in plain language. Crucially, this bridge can lower barriers for non-quantum specialists while accelerating exploration of quantum ready formulations.

We propose an LLM-guided pipeline that translates CML implementations into QML variants. Starting from user-provided core CML code for the model, the system (1) analyzes the model structure and recommends an appropriate QML paradigm based on hardware and noise constraints, (2) automatically generates candidate encodings, ansatz topologies, and measurement heads, (3) compiles backend-aware circuits, and (4) outputs executable QML code with inline documentation for non-quantum specialists. This work aims to achieve two goals: improving accessibility by lowering the entry barrier for ML practitioners to experiment with QML, and enhancing efficiency by reusing established CML design patterns while highlighting quantum-specific trade-offs such as circuit depth, sampling cost, and trainability. We evaluate the approach across representative tasks, analyzing fidelity to specification and failure modes, and position LLMs as a practical bridge between CML workflows and quantum-native implementations \citep{dupuis2024qiskit,cerezo2021variational}.

\section{Related Work}

LLM assisted quantum software is an active area spanning assistant style tooling and benchmarking. IBM's Qiskit Code Assistant combines a domain tuned LLM with Qiskit idioms and patterns for interactive code generation in VS Code and Jupyter, reporting substantial improvements on quantum tasks in the Qiskit HumanEval benchmark \citep{dupuis2024qiskit}. Beyond synthesis, LLMs have been evaluated as explanatory agents for quantum algorithms, indicating that careful prompt design can yield useful step by step narratives of circuit behavior \citep{aloisiod2024exploring}. LLMs can assist VQA workflows across (a) ansatz design (\eg, proposing hardware-efficient entanglers or problem inspired structures), (b) training scaffolds (gradients, parameter shift templates), and (c) experiment control (measurement grouping, shot allocation). Although peer reviewed evidence here is limited, the broader VQA literature provides the technical substrate LLMs must respect, including barren plateau aware cost design and general parameter shift rules \citep{cerezo2021variational, schuld2021supervised}. For QML specific tasks, LLMs have been applied to: (i) generate QML boilerplate and unit tests (datasets, encoders, training/evaluation loops) aligned to SDK idioms \citep{dupuis2024qiskit}, (ii) map classical ML abstractions (losses, model blocks) into QML analogues (feature maps, PQCs, measurement heads), and (iii) produce natural language explanations for quantum kernels and variational classifiers \citep{aloisiod2024exploring}. Our contribution complements this line by translating entire classical ML programs (\eg, scikit-learn or PyTorch code) into runnable QML variants, with design rationales surfaced inline to support non-quantum users.

\subsection{LLMs for Quantum Computing}

Large language models are increasingly used to assist quantum software tasks that are traditionally difficult for non experts. Early systems focus on code synthesis and evaluation for mainstream quantum SDKs. IBM's domain tuned Qiskit Code Assistant fine tunes a code LLM on Qiskit corpora and introduces an execution-based benchmark (Qiskit HumanEval) to measure quantum specific code generation quality, showing sizable gains over general-purpose code models \citep{dupuis2024qiskit, qiskit_human_eval}. In parallel, Guo et al. \citep{guo2024repairing} study program repair using ChatGPT to patch real bugs in the Bugs4Q benchmark, successfully repairing $29/38$ cases under a semi-automated workflow and highlighting typical failure modes (\eg, semantic misunderstandings of measurement and state reset). Easttom \citep{easttom2024utilizing} reports smaller-scale evidence that LLMs can propose algorithmic scaffolding and parameter update hints for standard quantum algorithms, while cautioning about correctness verification on hardware or faithful simulators.

Beyond SDK level assistance, several works target circuit synthesis and optimization for variational workloads. The Agent-Q framework fine tunes an LLM to emit pure quantum circuits and evaluates outputs by (i) syntactic validity, (ii) closeness of expectation values to optimized targets, and (iii) similarity of full outcome distributions metrics that align with VQE/QAOA practice \citep{jern2025agent}. These results support using LLMs to propose hardware aware ans\"atze or warm starts for VQAs in the NISQ regime. Complementing these application level studies, Melko and Carrasquilla argue that language model architectures (autoregressive/attention) are well suited to learning quantum state distributions and device behaviors, providing a conceptual bridge between token sequences and qubit strings \citep{melko2024language}. Zhou et~al.\ push this empirically by training LLMs to simulate small qubit circuit evolutions and predict output states efficiently, suggesting an LM-based surrogate for certain quantum dynamics \citep{zhou2025application}. Finally, work from the broader ML community explores quantum inspired operators for LLM training (\eg, high order products) that improve stability and compatibility in large scale optimization, while not operating on quantum hardware, this cross fertilization informs how quantum principles might shape future AI tooling for QC \citep{xiong2025reinvent}.

\subsection{LLMs for Code Generation}

For deep-learning practitioners, QML preserves familiar structure feature maps, layers, losses, optimizers, but replaces linear layers with unitaries and nonlinearities with measurement induced statistics. Hardware depth and connectivity become the analogs of parameter budgets and latency, while trainability and sampling cost, and throughput dominate performance. Our LLM based translation pipeline makes these correspondences concrete, reducing onboarding friction and accelerating QML exploration.

Code Generation tasks can be further broadly categorized into three tasks: \textbf{Code-to-Code (C2C)}, \textbf{Code-to-Natural Language (C2NL)}, and \textbf{Natural Language-to-Code (NL2C)} \citep{jiang2024survey}.

\textbf{Code-to-Code}. The C2C \citep{gong2024ast, zhang2025function, ibrahimzada2025alphatrans, li2024few, joos2025codecureagent} category focuses on generating code from existing code inputs such as partial contexts or complete programs. It comprises three core tasks: \textit{Code Completion}, \textit{Code Translation}, and \textit{Code Modification}. Code Completion predicts and suggests subsequent code segments based on the current context to enhance development efficiency. Code Translation converts source code between programming languages while preserving functionality and logic. Code Modification involves automatically detecting and correcting bugs to improve reliability, generating mutated code variants to assess robustness, or producing new test cases to evaluate performance and correctness. \textbf{Code-to-Natural Language}. The C2NL \citep{khan2022automatic, wen2022code2tree, makharev2025code, hong2025retrieval} paradigm transforms code into human-readable text. It focuses on producing concise textual explanations or documentation of code to enhance comprehension and maintenance. \textbf{Natural Language-to-Code}. The NL2C \citep{feng2020codebert, yu2024humaneval, sirovs2024github, li2023starcoder} paradigm represents the inverse direction, converting natural-language specifications into executable code. This task aims to generate functional source code from textual problem descriptions or instructions, streamlining development workflows and reducing manual coding effort. It stands as the central focus of LLM-based approaches for bridging human intent with automated code synthesis.

Q-Bridge, a code translation approach within the C2C category, leverages the strong mapping capabilities of LLMs to enable seamless transformation from classical AI to quantum computing, effectively bridging the gap between CML and QML.

\section{Methodology}

\begin{figure}[H]
    \vspace{-4mm}
    \centering
    \includegraphics[width=1.0\linewidth]{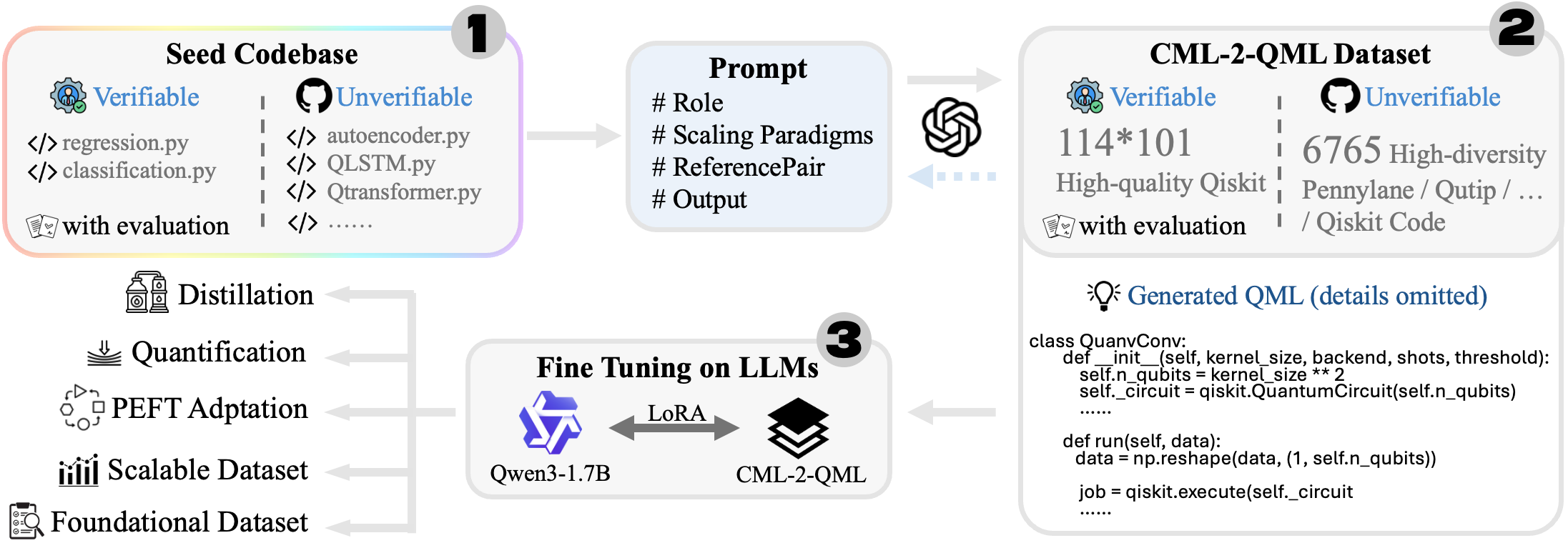}
    \vspace{-3mm}
    \caption{\textbf{The building of Q-Bridge.} It begins with \ding{182} establishing a robust seed codebase to form a reliable high-quality prototype, followed by \ding{183} employing LLMs to expand and refine the dataset, and finally \ding{184} training LLMs on the CML-2-QML dataset to obtain Q-Bridge model. Finally, we outline potential future directions of our work.} 
    \label{fig:2}
    \vspace{-4mm}
\end{figure}

High-quality data is the foundation of a powerful code translation model. The performance and adaptability of LLMs depend directly on the scale, diversity, and precision of their training data \citep{nijkamp2022codegen,chen2021evaluating,wang2023codet5+}. Rich, domain-specific datasets enable models to learn deeper structural and semantic mappings, while poor or limited data lead to weak generalization. Continuous data curation is therefore essential for building a robust \our model.

\our follows a self-involving pipeline \citep{haluptzok2022language} that iteratively expands and refines the dataset. We first construct a high-quality seed codebase composed of two parts: \textbf{(1).} verifiable Qiskit code written by human experts and equipped with corresponding evaluation tools, and \textbf{(2).} diverse quantum code collected from open-source communities such as GitHub, covering various quantum libraries and model architectures, all reviewed by domain experts. Using tailored prompts \citep{zeng2024visual, zeng2025mept}, LLMs are then employed to scale up the seed codebase through extension, combination, and controlled modification, producing a 6K-scale unverifiable CML-2-QML dataset and a verified subset containing 114 ansatze and 101 feature maps to ensure both high quality and high diversity. This CML-2-QML dataset is used to fine-tune LLMs via supervised low-rank adaptation (LoRA) to obtain the \our model. The scaling stage is repeated iteratively, allowing the model to evolve through self-involvement. Finally, our future direction focuses on advancing scalable quantum AI through distillation, quantification, parameter-efficient fine-tuning (PEFT) adaptation, and the continuous expansion of foundational datasets.

\subsection{Stage1: Foundation of Seed Codebase}

We begin by constructing a robust seed codebase that serves as the foundation for the entire self-involving pipeline. This stage focuses on curating, collecting, cleaning, and validating high-quality quantum code to ensure a reliable starting point for large-scale dataset generation and model training. 

\subsubsection{Verifiable Data Curation}

This process involves curating verifiable Qiskit code written by human experts. Each script is executable and paired with evaluation tools to ensure functional correctness. Through strict verification, we retain only reproducible and well-structured examples that form the verified core of our quality codebase.

\subsubsection{Unverifiable Data Collection}

To increase diversity, we collect unverifiable quantum code from open-source repositories such as GitHub and official library documentation. These codes, covering multiple quantum frameworks and architectures (\eg, TorchQuantum, Qutip, and Pennylane), are manually reviewed by experts to confirm their structural soundness and conceptual relevance. Through strict cleaning, we remove incomplete, redundant, or logically inconsistent code, leaving well-structured examples that form our diveristy codecase. Although not directly testable, they enrich the dataset’s coverage of quantum modeling patterns.

\begin{tcolorbox}[colback=lightgray!10, colframe=black,  title={Part 1: Scaling Prompt for Single-File}]
\begin{Verbatim}[breaklines=true, breaksymbolleft={}, fontsize=\footnotesize]
#Role
You are gpt-oss-20b, an expert hybrid quantum-classical ML engineer who upgrades seed projects into robust research assets.

#Goal
Given a single reference pair, craft a substantially more capable ML/QML duo that honours the original while moving it forward.

#Inputs
- `#TargetOutput` block describing the intended file location, anchor seed, and reference count.
- One `#ReferencePair[1]` block providing the matched ML/QML seed with paths and fenced code excerpts.
- Optional repository index notes.

#Output Contract
Always populate the structured output fields with self-consistent content using the exact template below. The literal tokens (including `assistantfinal`) and triple quotes are required so downstream tooling can parse the generation without ambiguity:

```
assistantfinal
name: <SharedClassName>
scaling_paradigm: <extension|controlled modification>
summary: <2-3 short sentences on the upgrade>
ml_code: '''
<importable Python module that defines SharedClassName>
'''
qml_code: '''
<importable quantum Python module that defines SharedClassName>
'''
```

(continued below)
\end{Verbatim}
\end{tcolorbox}

\begin{tcolorbox}[colback=lightgray!10, colframe=black,  title={Part 2: Scaling Prompt for Single-File}]
\begin{Verbatim}[breaklines=true, breaksymbolleft={}, fontsize=\footnotesize]
(continued from above)

Do not emit any additional commentary, Markdown fences, or fields before, between, or after the entries in the template.

#Scaling Paradigms
1. **extension** — deepen the seed by adding richer capabilities, pipelines, or experiments.
2. **controlled modification** — introduce a deliberate variation that reframes part of the anchor while keeping it recognisable.

#Constraints
- Keep the ML output entirely classical (NumPy, PyTorch, scikit-learn, etc.).
- Ensure the QML output features a quantum-centric contribution (variational circuits, simulators, qiskit/pennylane/Braket APIs, etc.).
- Do not copy seeds verbatim; show original yet traceable evolution.
- Produce self-consistent files with fixed imports, names, and signatures.
- Focus on the modification of model structure rather than the training pipelines.

#Generation Checklist
1. Both outputs meaningfully diverge from their seed yet remain compatible.
2. Comments and docstrings clarify why the scaling choices matter.
3. ML/QML components expose complementary experiments or capabilities inspired by the reference.

#Tone
Be authoritative, succinct, and technically precise. Use bullets and short paragraphs.

#Automation Template
Automation adds context in this shape:
```
#TargetOutput
relative_path: <target_relative_path>
anchor_reference: <anchor_relative_path>
reference_pair_count: <N>

```python
#ReferencePair[1]
relative_path: <seed_relative_path>
ml_seed_path: seed_codebase/ML-Github/<seed_relative_path>
qml_seed_path: seed_codebase/QML-Github/<seed_relative_path>

```python
# ML seed code...
```

```python
# QML seed code...
```
```
Treat the block as authoritative context for the transformation.
\end{Verbatim}
\end{tcolorbox}

\begin{tcolorbox}[colback=lightgray!10, colframe=black,  title={Paradigm Prompt in Scaling Prompt for Multi-File}]
\begin{Verbatim}[breaklines=true, breaksymbolleft={}, fontsize=\footnotesize]
#Combination Expectations
- Integrate substantive ideas from each reference pair instead of favouring a single seed.
- Highlight how the classical and quantum halves reinforce the merged concept.
- Preserve compatibility with the anchor paths while expanding capabilities.
\end{Verbatim}
\end{tcolorbox}

\subsection{Stage2: Scaling Up the Dataset}

\begin{wrapfigure}{r}{0.6\textwidth}
\vspace{-4mm}
\includegraphics[width=0.6\textwidth]{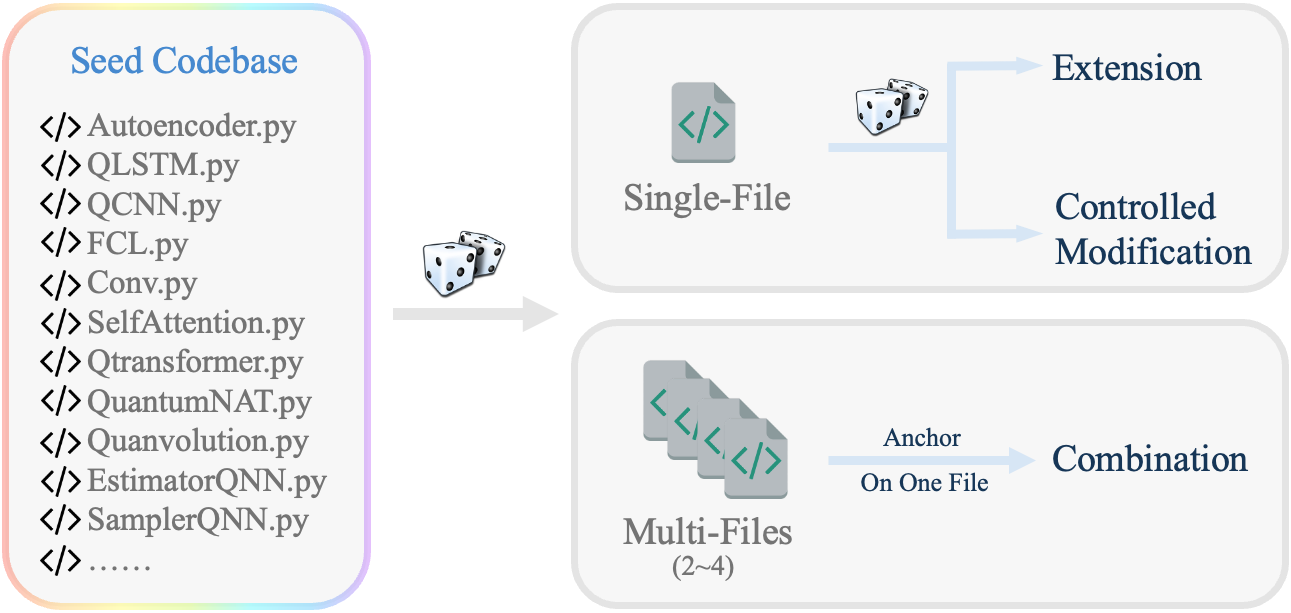}
\vspace{-2mm}
\caption{The pipeline for scaling from the seed codebase.}
\label{fig:3}
\vspace{-2mm}
\end{wrapfigure}

Using the curated seed codebase, we employ LLMs (\ie, gpt-oss-20b \citep{agarwal2025gpt}) with carefully designed prompts as shown above to scale up the dataset. The models perform controlled operations such as code expansion, combination, and small-scale mutation to generate new, high-quality quantum code. Through iterative generation and self-verification, we build a 6K-scale dataset that balances quality and diversity, forming the core training corpus for instruction tuning.

As shown in Figure \ref{fig:3}, for each scaling iteration on the unverifiable codes, we first assign weighted probabilities to select 1–4 reference files. For single-file references, the model randomly applies either the extension or controlled modification paradigm to generate ansatz and feature map variants, while applying only extension for CML–QML pairings. For multi-file references, one file is chosen as the anchor, while complementary insights from the remaining references are integrated to construct more complex and semantically enriched code. After scaling, we apply a syntax checker to ensure all generated code is grammatically correct.

\subsection{Stage3: Instruction Tuning on LLMs}

The scaled dataset is then used to perform supervised fine-tuning (SFT) on large language models with LoRA, enabling efficient parameter updates while preserving the base model’s generalization. The fine-tuning process emphasizes code consistency, correctness, and explainability. This stage outputs the Q-Bridge, a model capable of generating accurate and interpretable quantum code.

Formally, SFT process minimizes a token-level loss between the model’s predictions and ground-truth quantum code tokens:

\begin{equation}
\mathcal{L}_{\text{SFT}} = - \sum_{t=1}^{T} \\log P_{\theta}(y_t \mid y_{<t}, x),
\end{equation}

where $x$ denotes the CML input, $ y_t $ the $t$-th target token in the corresponding QML code, and $ P_{\theta} $ the conditional probability defined by the model parameters $\theta$. This objective ensures the model aligns syntactic and semantic structures between the classical and quantum domains.

Specifically, the LoRA-augmented SFT process minimizes a token-level loss while constraining updates through low-rank matrices:

\begin{equation}
\mathcal{L}_{\text{LoRA-SFT}} = - \sum_{t=1}^{T} \log P_{\theta + \Delta W}(y_t \mid y_{<t}, x), \quad \text{where } \Delta W = A B, \; A \in \mathbb{R}^{d \times r}, \; B \in \mathbb{R}^{r \times k}.
\end{equation}

Here, $\Delta W$ represents the low-rank adaptation matrices that efficiently fine-tune the model without full parameter updates, enabling scalable and memory-efficient training for large-scale instruction tuning.

\subsection{Future Direction}

After crafting the CML-2-QML dataset and Q-Bridge model, we build a foundational model for code translation from CML to QML, bridging the gap for the development of quantum AI. Furthermore, we plan to facilitate PEFT for future updates of quantum-related libraries, thereby alleviating the retraining burden, and employ model distillation and quantification to enable lightweight deployment. Regarding dataset scalability, since the performance of LLMs largely depends on the quality of their training data, our dataset represents the first large-scale, high-quality collection of paired CML and QML code, serving as a scalable and foundational resource for building more powerful baselines in the future. The discussion guides the next iteration of the self-involving pipeline, completing the feedback loop of continuous improvement.

\subsubsection{PEFT Adptation}

One of the major challenges in QML code translation lies in the rapidly evolving library APIs. Unlike classical ML, where frameworks such as PyTorch maintain relatively stable and backward-compatible APIs, QML libraries are updated frequently, often introducing breaking changes within a few months. This instability raises concerns about model adaptability, as translated code may become incompatible with newer library versions. To address this, we plan to adopt PEFT adaptation \citep{zeng2026tokenseek}. PEFT freezes most model parameters and updates only a small subset using newly collected datasets. This approach allows the model to retain its core translation capability while incrementally learning API changes. It efficiently updates partial knowledge without retraining from scratch, preventing catastrophic forgetting and ensuring consistent translation performance across evolving QML frameworks.

\subsubsection{Distillation}

We will employ knowledge distillation to transfer the capabilities of the large Q-Bridge model into smaller student models while maintaining comparable performance. The distilled variants (\eg, Llama-3.2-1B and Qwen2.5-0.5B) will learn from the teacher model’s intermediate representations and final, enabling efficient compression \citep{zeng2025probabilistic} without sacrificing accuracy in CML-to-QML translation. This approach reduces computational overhead during training and inference, making the model more adaptable for real-world applications with limited hardware resources. In addition, iterative distillation cycles will be used to refine the student models’ generalization ability, ensuring stability and consistency across different quantum programming frameworks. Through this process, we aim to establish a family of lightweight yet high-fidelity Q-Bridge models suitable for scalable deployment and continual updates.

\subsubsection{Quantification}

We will incorporate quantification to further reduce model size and inference latency while maintaining numerical stability and performance. By applying mixed-precision and low-bit quantization techniques (\eg, 8-bit and 4-bit quantization), the Q-Bridge model can achieve efficient memory utilization and faster computation without significant degradation in translation quality. This process enables the deployment of the distilled Q-Bridge variants on resource-constrained environments, such as edge devices or quantum simulators with limited GPU availability. Moreover, quantization-aware fine-tuning will be adopted to ensure robustness during precision reduction, facilitating seamless integration into PEFT workflows. This optimization step will complete the lightweight adaptation phase of our pipeline, making the CML-2-QML translation system both scalable and hardware-efficient.

\subsubsection{Scalable Dataset}

Compared to large-scale datasets such as Wikipedia and The Pile, which contain millions of samples, the CML-2-QML dataset is of a smaller scale, comprising thousands of examples. This limitation primarily reflects the current stage of the quantum computing community, which is still in development and lacks abundant high-quality open-source QML resources. However, due to our evolving pipeline design, the dataset can be further expanded through the scaling prompts introduced above as more reliable QML code becomes available within the community, highlighting the strong potential of the CML-2-QML dataset for future scalability.

\subsubsection{Foundational Dataset}

In this paper, we present a standardized and executable training pipeline and accompanying evaluation suite for constructing the CML-2-QML dataset. This pipeline ensures full reproducibility across data preprocessing, model execution, and validation, enabling consistent benchmarking for code translation tasks. Each curated sample is verified for syntactic integrity and functional correctness through automated testing and expert review. By maintaining unified data formats and fixed evaluation scripts, the foundational dataset provides a reliable ground truth for scaling, instruction tuning, and downstream QML benchmarking.

\section{Study}

\subsection{Study of CML-2-QML Dataset}

\subsubsection{Analysis of Verifiable Seed Codebase}

\paragraph{Overview}

Two dominant QML paradigms are (1) quantum enhanced classical ML, which leverages quantum subroutines for kernel evaluation or linear algebra, and (2) native PQC models are trained in an end to end way. Quantum kernels embed classical data into quantum states via feature maps and estimate inner products $\langle \phi(x),\phi(x')\rangle$ on hardware \citep{havlivcek2019supervised,schuld2021supervised}. Variational models employ trainable unitaries $U(\theta)$ with measurements defining task specific objectives \citep{perez2020data,cong2019quantum}. Encoding strategies (angle, amplitude, basis/Pauli-feature maps) control expressivity and measurement complexity. Data re-uploading interleaving feature encodings with trainable layers achieves universal approximation with few qubits and shallow depth \citep{perez2020data}. Circuit capacity and inductive bias depend on entangling structure and depth, with expressibility and entangling capability metrics guiding ansatz selection \citep{sim2019expressibility}. Hardware compatible differentiation uses parameter shift and related rules to compute exact gradients from expectation values \citep{schuld2021supervised}. These enable gradient based optimizers for PQCs without full tomography. Measurement cost, shot noise, and readout error drive estimator variance and training time. 

Barren plateaus (exponentially vanishing gradients) arise from certain random or global-cost landscapes; cost design and locality can mitigate scaling issues \citep{mcclean2018barren, cerezo2021variational}. Architectural choices, initialization, and problem inspired ansätze further affect trainability. QML generalization has formal bounds that depend on the number of trainable gates and the encoding strategy \citep{caro2023out,caro2021encoding}. Notably, the power of data results show that when learning from examples, classical models can compete with, or even surpass, quantum models on some tasks, underscoring the need to identify regimes where quantum structure truly matters \citep{huang2021power}. These insights motivate careful problem selection and benchmarking when translating CML workflows into QML. For deep learning native readers, QML mirrors familiar patterns (feature maps, layers/blocks, losses, optimizers) but substitutes linear layers with unitaries and nonlinearities with measurement induced statistics. Depth/connectivity constraints play the role of parameter budgets and latency constraints in classical systems, while trainability and sampling cost replace flops throughput as primary performance bottlenecks.

\begin{table}
\small
\setlength{\tabcolsep}{4pt}
\renewcommand{\arraystretch}{1.12}
\begin{tabularx}{\textwidth}{|p{3.4cm}|X|X|X|X|}
\hline
\multicolumn{5}{|c|}{\textbf{QML Model, Description, Applications, Ansatz, and Feature Map}}\\ 
\hline
\thead{QML Model} & \thead{Description} & \thead{Applications} & \thead{Ansatz} & \thead{Feature Map} \\ 
\hline
Continuous-Variable QNN \citep{killoran2019continuous} &
Layered QNN using Gaussian (affine) + non-Gaussian (activation) gates &
Fraud detection; autoencoding &
Interferometers + squeezers + displacements + non-Gaussian gates &
Displacement encoding of input data on vacuum modes \\ 
\hline
Quanvolutional NN \citep{henderson2020quanvolutional} &
Local quantum circuits on image patches to augment CNNs &
Image recognition/classification &
Random small quantum circuit per patch &
Binary threshold encoding \\ 
\hline
Deep QNN \citep{beer2020training} &
Parametrized unitaries with fidelity-based training; depth-robust &
Learning an unknown unitary &
Stacked “quantum perceptron” unitaries on qubits &
Direct quantum input (\eg |0…0⟩); no separate classical feature map \\ 
\hline
Quantum Autoencoder \citep{bondarenko2020quantum} &
Compress/denoise entangled states with an autoencoder circuit &
Restore clean states; metrology &
Feedforward QNN (quantum perceptron layers) &
Quantum input states directly (no classical encoding) \\ 
\hline
Quantum GNN \citep{beer2020training} &
QML for graph structured quantum data with self-supervised loss &
Graph-data learning; noise-aware BO &
Universal QNN ansatz (arbitrary PQC on all graph vertices) &
Quantum input states at vertices (encoded by graph edges), no classical mapping \\ 
\hline
Quantum Transformer \citep{wang2022torchquantum} &
Attention blocks + PQC modules in a hybrid model &
Quantum feature encoding &
PQC token-embedding blocks (parameterized rotations forming a unitary) &
Classical embeddings encoded via a linear layer + angle rotations into PQCs \\ 
\hline
Quantum LSTM \citep{wang2022torchquantum} &
Hybrid LSTM with PQC-based gates for sequences &
NLP; temporal modelling &
PQC-based quantum gates (trainable rotation/entangler layers in each LSTM gate) &
Sequential input encoded by rotation gates (angle encoding into qubits) \\ 
\hline

\end{tabularx}
\caption{Quantum machine learning models, descriptions, applications, ansatze, and feature maps.}
\label{table:qml_models_raw1}
\end{table}

\begin{table}
\small
\setlength{\tabcolsep}{4pt}
\renewcommand{\arraystretch}{1.12}
\begin{tabularx}{\textwidth}{|p{3.4cm}|X|X|X|X|}
\hline
\multicolumn{5}{|c|}{\textbf{QML Model, Description, Applications, Ansatz, and Feature Map}}\\ 
\hline
\thead{QML Model} & \thead{Description} & \thead{Applications} & \thead{Ansatz} & \thead{Feature Map} \\ 
\hline
Quantum SVM (Kernel) \citep{wang2022torchquantum} &
Compute quantum kernel via circuit; train classical SVM &
Separability and classification &
Kernel PQC (\eg ZZ-feature map circuit with entanglement) &
Angle embedding of features (\eg Z-feature map) \\ 
\hline
Quanvolution (TorchQuantum) \citep{wang2022torchquantum} &
Quantum circuits as convolution-like filters &
Image recognition and classification &
Fixed PQC filters (random or structured rotations+CX on patches) &
Patch pixels encoded into qubit angles (\eg via RY gates) \\ 
\hline
QuantumNAT \citep{quantumNAT} &
Noise-aware PQC training (injection, normalization, quantization) &
Classification on NISQ devices &
Layered QNN (PQC classifier) with trainable weights &
Classical inputs encoded via encoder layer (angle rotations) \\ 
\hline
QNN Regression \citep{wang2022torchquantum} &
PQC for predicting continuous outputs &
Regression and prediction &
PQC similar to classification QNN (rotation+entangler layers) &
Angle encoding of input features \\ 
\hline
QCNN \citep{cong2019quantum,vatan2004optimal} &
Quantum convolution + pooling layers for patterns &
Image classification &
Parametrized two-qubit unitaries (convolution filters); pooling discards qubits &
ZFeatureMap (phase rotation map on input pixels) \\ 
\hline
Fast Prototyping QNN \citep{10821318} &
Qiskit–PyTorch module; up to $100\times$ faster VQA training &
Classification and reinforcement learning &
Data-reuploading ansatz: layers of RX/RY/RZ + nearest-neighbor CNOTs &
Data re-uploading: repeated angle encoding with trainable scaling \\ 
\hline
Estimator/Sampler QNN \citep{qiskit_ml_tutorial_qnn} &
Build/train VQMs in hybrid workflows &
Integrate QNNs with classical ML &
Example circuits: 1-qubit H–RY–RX (estimator), 2-qubit RY–RY–CX–RY (sampler) &
Angle encoding via RY rotations of input values \\ 
\hline
\end{tabularx}
\caption{Quantum machine learning models, descriptions, applications, ansatze, and feature maps similar to shown in \autoref{table:qml_models_raw1}.}
\label{table:qml_models_raw2}
\end{table}

In our work, we employ a \emph{feature map} to embed classical feature vectors into quantum states, enabling downstream processing on a quantum circuit. We also use a parameterized \emph{ansatz}—composed primarily of rotation gates—whose parameters are updated over training epochs to minimize a task specific loss. This joint design (data embedding  trainable circuit) drives the quantum model toward lower prediction error during QML training. The QML models summarized in \autoref{table:qml_models_raw1} and \autoref{table:qml_models_raw2} each employ their own variations of the \emph{feature map} and \emph{ansatz} (\eg, encoding order, entanglement topology, repetition depth). These choices are tailored to the task and hardware constraints of each model. For a deeper understanding of the specific circuits, including their parameterizations and design rationale, please consult the per-model descriptions associated with \autoref{table:qml_models_raw1} and \autoref{table:qml_models_raw2}.

\textbf{Feature map (ZZFeatureMap).} 
We embed an $n$-dimensional real vector $\mathbf{x}\in\mathbb{R}^n$ into an $n$-qubit state with Qiskit’s \texttt{ZZFeatureMap}, a second-order Pauli-$Z$ evolution circuit. The construction prepares $\lvert +\rangle^{\otimes n}$ and then, for a chosen entanglement graph $E$ and number of repetitions $r$, applies data-dependent single-qubit $Z$ phases and pairwise $ZZ$ couplings:
\[
U_{\mathrm{ZZ}}(\mathbf{x})
=\Bigg[\;\prod_{j=1}^{n} e^{\,i\,\alpha\,\varphi(\,x_j\,) Z_j}\;\;\prod_{(i,j)\in E} e^{\,i\,\alpha\,\varphi(\,x_i,x_j\,)\, Z_i\!\otimes\! Z_j}\,\Bigg]^{\!r}.
\]
By default, the scale is $\alpha=2$, the pointwise map is $\varphi(x)=x$, and the pairwise map is $\varphi(x_i,x_j)=(\pi-x_i)(\pi-x_j)$, while $E$ can be chosen (\eg, \texttt{linear}, \texttt{full}). This realizes a second order feature map that captures both unary and pairwise interactions of the input features via commuting $Z$/$ZZ$ evolutions.

\textbf{Ansatz (RealAmplitudes).}
For the variational layer we use Qiskit’s \texttt{RealAmplitudes}, which alternates parameterized $R_y$ rotations with layers of CNOT entanglers according to a user selectable connectivity pattern $E$ (\eg, \texttt{linear}, \texttt{reverse\_linear}, \texttt{full}), repeated $r$ times. Writing the parameters at repetition $\ell$ as $\boldsymbol{\theta}^{(\ell)}\in\mathbb{R}^n$, the ansatz factors as
\[
U_{\mathrm{RA}}(\boldsymbol{\theta}) \;=\; \Bigg[\; U_{\mathrm{ent}}(E)\;\cdot\; \bigotimes_{q=1}^{n} R_y\!\big(\theta^{(\ell)}_q\big)\;\Bigg]^{\!r}
\quad,
\]
which prepares states with purely real amplitudes (hence the name) while offering a simple, hardware efficient pattern of local rotations plus CX entanglement.

In our pipeline for learning from raw datasets, we use a mix of quantum models as displayed in \autoref{table:qml_models_raw1} and \autoref{table:qml_models_raw2} to (i) denoise/compress inputs before tokenization, (ii) extract robust features for downstream LLM objectives, and (iii) stress test training under distribution shift and noise. Quantum autoencoders compress and denoise state encodings of data blocks, preserving salient structure while discarding components useful as a prefilter for low trust corpora. Kernel based QML (quantum SVM/quantum kernels) maps examples into high dimensional Hilbert spaces where linear boundaries approximate complex semantics; we use this both for robust filtering (accept/reject) and for small, high precision calibration tasks. Quanvolution neural networks apply quantum convolutions and pooling to patches (text, image, or mixed tokens after embedding), giving noise tolerant local features that improve early screening and multimodal grounding. Deep QNNs / CV-QNNs (Gaussian + non-Gaussian layers) serve as compact function approximators for auxiliary losses (\eg, label correction, quality scoring) where we need expressive, low parameter models. To handle structure beyond independent and identically distributed samples, quantum GNNs exploit graph relations (document links, citation graphs, conversational threads) to regularize weak labels. 

For sequence tasks, quantum LSTM and quantum augmented transformers swap selected gates with PQCs, probing whether quantum non-linearities improve long context robustness and out of distribution generalization. Finally, QuantumNAT style noise aware training (injection, normalization, quantization) is used across these modules to close the sim to hardware gap and harden models against label and sensor noise. Operationally, we keep everything hybrid and production friendly. EstimatorQNN and SamplerQNN integrate with PyTorch for differentiable training alongside the LLM, and our Qiskit–PyTorch acceleration path gives orders of magnitude speedups for inner loops on a single machine. In practice quantum autoencoders and kernels front load data cleaning and triage, \ie, QCNNs provide local, noise robust features. QGNNs inject relational priors, quantum LSTM blocks explore sequence level gains, and QuantumNAT hardens the whole stack. Together, these modules improve signal to noise, calibration, and robust generalization when training LLM components on imperfect, partially verified data.

\paragraph{Evaluation Framework}

An evaluation framework is established to assess both code quality and functionality. It includes automated test scripts, metric-based scoring for execution correctness, and expert feedback loops for semantic evaluation. This framework provides a quantitative and qualitative foundation for later stages of model tuning and dataset scaling.

We evaluate on five datasets (\ie, \emph{Iris-Regression-3f}, \emph{Diabetes-10f}, \emph{CaliforniaHousing-8f}, \emph{Synthetic-4f}, \emph{Synthetic-6f}) using community splits when available or a stratified 80/10/10 train/validation/test partition with class balance and no leakage.

For verified code for regression, we evaluate models with $k$-fold cross validation and report, \ie, average mse, the mean of the per fold mean squared error,  $\mathrm{MSE}=\frac{1}{n}\sum_{i=1}^{n}\bigl(y_i-\hat{y}_i\bigr)^2$, and stdandard mse, the standard deviation of those per fold mse (smaller implies more stable performance across folds). We also track residual diagnostics, \ie, residual mean$\frac{1}{N}\sum_{i=1}^{N}\bigl(y_i-\hat{y}_i\bigr)$, which should be near zero for an unbiased model (on a standardized target, values within $\pm 0.05$ are typically fine), and residual stdandard deviation, the standard deviation of residuals that captures the spread of errors. By default the target is standardized target on original scale is set to false, so avg mse is in standard deviation squared units (\eg, $\text{avg\_mse}=1.0 \Rightarrow \mathrm{RMSE}\approx 1$ target s.d.; $\text{avg\_mse}=0.25 \Rightarrow \mathrm{RMSE}\approx 0.5$ s.d.). If \texttt{target\_on\_original\_scale=true}, all error metrics are reported in the original units (\eg, dollars, mm). Finally, number of residuals is the total number of residuals accumulated across validation folds (usually equal to the dataset size when each sample appears exactly once in validation).

For classification, we likewise use k-fold cross-validation and summarize avgerage accuracy (mean fraction correct across folds; higher is better) with standard accuracy (stability across folds), and average $F_1$ (mean of per class $F_1$ scores, unweighted; higher is better) with standard $F_1$ macro. Accuracy gives an overall hit rate, while macro $F_1$ is robust to class imbalance by weighting all classes equally. When relevant, we additionally report AUROC (area under the ROC curve and threshold free ranking quality), $log$ $loss$ cross entropy (penalizes overconfident wrong predictions, \ie, lower is better), and the $Brier$ $score$ (mean squared error of predicted probabilities; lower is better) to assess calibration. Together, these metrics capture correctness (accuracy), balance across classes (macro $F_1$), ranking quality (AUROC), and probabilistic calibration ($log-loss$/$Brier$), plus their across fold variability for reliability.

\paragraph{Evaluation Result}

\begin{table*}[h]
\vspace{-2mm}
\caption{Regression performance comparison between classical and quantum models in \emph{Iris-Regression-3f}.}
\vspace{-3mm}
\begin{adjustbox}{width=0.99\textwidth,center}
\begin{tabular}{l|c|cccc}
\toprule
\multirow{2}{*}{\textbf{Config}} & \multirow{2}{*}{\textbf{Optimizer}} & \multicolumn{4}{c}{\textbf{Regression Metrics}}  \\
 &  & \textbf{Avg MSE} & \textbf{Std MSE} & \textbf{Residual Mean} & \textbf{Residual Std} \\
\midrule
\midrule
\multicolumn{6}{c}{\textbf{Classical Machine Learning}}\\
\midrule
(Linear + ReLu) with hidden\_dims = [4, 16] & Adam & 0.9637 & 0.0926 & -0.1478 & 0.9705  \\
(Linear + ReLu) with hidden\_dims = [64, 32] & Adam & 0.8236 & 0.0713 & 0.0527 & 0.9060  \\
\midrule
\midrule
\multicolumn{6}{c}{\textbf{Quantum Machine Learning}}\\
\midrule
(ZZFeatureMap + RealAmplitudes) with num\_reps = 3 & COBYLA & 1.1074 & 0.1338 & -0.0031 & 1.0523 \\
\bottomrule
\end{tabular}
\end{adjustbox}
 \label{tab:3}
 \vspace{-2mm}
\end{table*}

Table~\ref{tab:3} presents the regression performance comparison between classical and quantum models on the \emph{Iris-Regression-3f} dataset. The classical neural networks demonstrate lower mean squared error (MSE) and more stable residuals compared to the quantum counterpart. Specifically, the model with hidden dimensions [64, 32] achieves the best performance, with an average MSE of 0.8236 and the smallest residual mean and variance, indicating effective convergence and low prediction bias. In contrast, the quantum model based on \texttt{ZZFeatureMap} and \texttt{RealAmplitudes} exhibits a higher average MSE (1.1074) and larger variability, suggesting that the current quantum circuit depth (\texttt{num\_reps = 3}) and the COBYLA optimizer may not fully capture the regression patterns. These results imply that while quantum models can approximate nonlinear relationships, their performance is still limited by circuit expressivity and optimizer efficiency relative to classical baselines. Overall, the feature matrix between classical and quantum models is aligned, indicating that direct code translation between the two paradigms is feasible.

\begin{table*}[h]
\vspace{-2mm}
\caption{Classification performance comparison between classical and quantum models in \emph{Iris-Regression-3f}.}
\vspace{-2mm}
\begin{adjustbox}{width=0.99\textwidth,center}
\begin{tabular}{p{7cm}|c|cccccc}
\toprule
\multirow{2}{*}{\textbf{Config}} & \multirow{2}{*}{\textbf{Optimizer}} & \multicolumn{2}{c}{\textbf{Accuracy}} & \multicolumn{2}{c}{\textbf{F1 Macro}} & \multicolumn{2}{c}{\textbf{F1 Weighted}} \\
 &  & \textbf{Mean} & \textbf{Std} & \textbf{Mean} & \textbf{Std} & \textbf{Mean} & \textbf{Std} \\
\midrule
\midrule
\multicolumn{8}{c}{\textbf{Classical Machine Learning}}\\
\midrule
\makecell[l]{(Linear + ReLu) with hidden\_dims = [8, 8]} & Adam & 0.340 & 0.071 & 0.168 & 0.027 & 0.177 & 0.062 \\
\midrule
\midrule
\multicolumn{8}{c}{\textbf{Quantum Machine Learning}}\\
\midrule
\makecell[l]{(ZZFeatureMap + RealAmplitudes)\\ with num\_reps = 2} & COBYLA & 0.380 & 0.088 & 0.375 & 0.078 & 0.379 & 0.081 \\
\bottomrule
\end{tabular}
\end{adjustbox}
\label{tab:4}
\vspace{-2mm}
\end{table*}

Tables~\ref{tab:4} and~\ref{tab:confusion} report the classification performance and detailed per-class accuracy of both classical and quantum models on the \emph{Iris-Regression-3f} dataset. The quantum model slightly outperforms the classical baseline in terms of overall accuracy (0.380 vs. 0.340) and F1 scores across all metrics, suggesting that the quantum circuit is capable of capturing more discriminative features within this small dataset. Despite the performance gain, both models exhibit relatively low absolute accuracies, indicating that the dataset’s regression-style formulation or limited sample size constrains classification separability.

\begin{table*}[h]
\vspace{-2mm}
\caption{Per-class accuracy and confusion matrix for classification results.}
\vspace{-2mm}
\begin{adjustbox}{width=0.5\textwidth,center}
\begin{tabular}{l|ccc|ccc}
\toprule
 & \multicolumn{3}{c}{\textbf{CML}} & \multicolumn{3}{c}{\textbf{QML}} \\
\midrule
\textbf{Class} & I & r & i & I & r & i \\
\midrule
\textbf{Accuracy} & 0.64 & 0.38 & 0.0 & 0.32 & 0.40 & 0.42 \\
\midrule
\textbf{Confusion} & 32 & 18 & 0 & 16 & 13 & 21 \\
\textbf{Matrix} & 31 & 19 & 0 & 13 & 20 & 17 \\
\textbf{of CML and QML} & 27 & 23 & 0 & 11 & 18 & 21 \\
\bottomrule
\end{tabular}
\end{adjustbox}
\label{tab:confusion}
\vspace{-2mm}
\end{table*}

From the confusion matrix, the classical model (CML) performs better on class “I” but fails completely on class “i,” while the quantum model (QML) distributes predictions more evenly across all classes, with each class achieving a moderate accuracy around 0.32–0.42. This pattern implies that the quantum circuit, though less sharply tuned, achieves a more balanced classification behavior. Overall, these findings highlight that while QML models are still less mature than classical ones in stability and scalability, they show potential advantages in balanced representation learning and mitigating class bias.

\subsubsection{Analysis of Unverifiable Seed Codebase}

\begin{figure}[h]
    \centering
    \includegraphics[width=1.0\linewidth]{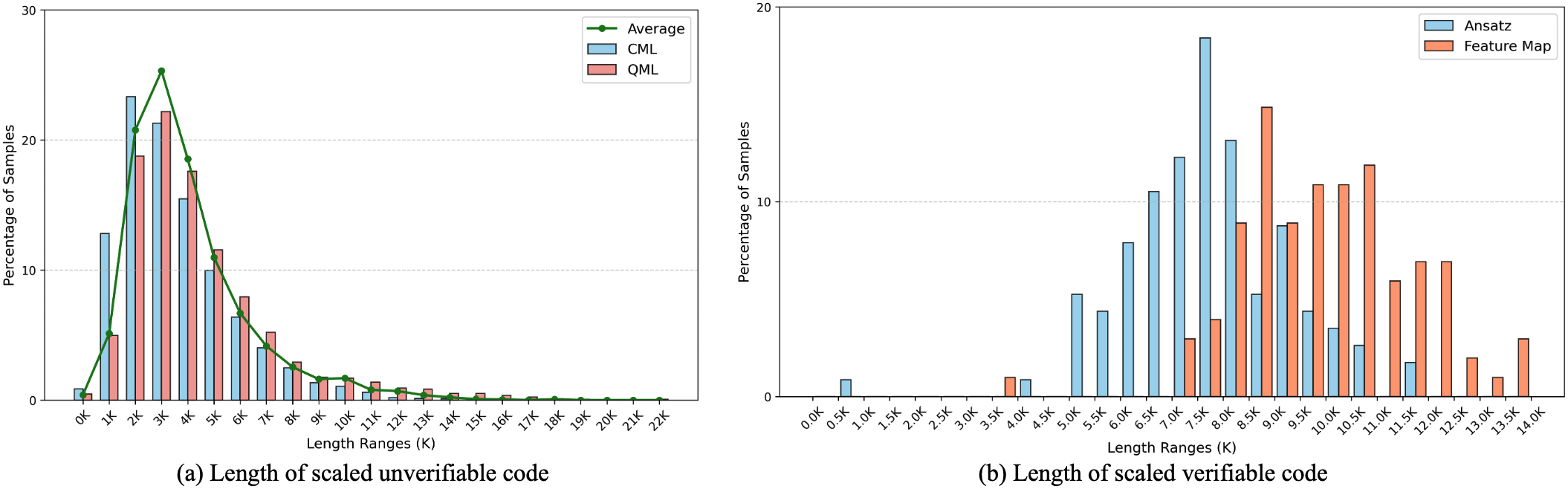}
    \vspace{-3mm}
    \caption{Distribution of scaled code lengths for classical and quantum codes.} 
    \label{fig:2}
    \vspace{-4mm}
\end{figure}

Figure~\ref{fig:2} illustrates the distribution of scaled code lengths across classical and quantum models. In Figure~\ref{fig:2} (a), the unverifiable code samples show a right-skewed distribution with most samples concentrated between 1\text{K} and 4\text{K} tokens, where both CML and QML exhibit similar trends. The QML code tends to be slightly longer on average, reflecting additional structural components such as circuit definitions and parameter bindings. The average curve indicates a sharp decline beyond 5\text{K} tokens, suggesting limited code complexity for most samples. In Figure~\ref{fig:2} (b), the verifiable code length distribution reveals that \texttt{Feature Map} components generally produce longer and more varied code than \texttt{Ansatz}, with peaks around 8\text{K}–10\text{K}. This implies that the verification stage emphasizes structural encoding and parameter mapping, leading to greater token expansion. Overall, the quantum code demonstrates higher variability and expressivity, while classical counterparts remain more compact and uniform.

Table~\ref{tab:6} summarizes the structural and quantitative characteristics of the seed codebase, including the average code length, reference counts, and paradigm distribution across classical and quantum implementations. Overall, the dataset spans 18 representative models, covering a diverse range of architectures from foundational neural networks (\ie, \texttt{Autoencoder}, \texttt{FCL}) to hybrid quantum structures (\ie, \texttt{QCNN}, \texttt{QLSTM}, \texttt{QTransformerTorch}). The average code length varies widely, from only a few hundred tokens (\ie, \texttt{EstimatorQNN}, \texttt{SamplerQNN}) to nearly ten thousand (\ie, \texttt{QTransformerTorch}), indicating substantial diversity in model complexity and implementation granularity. This variation reflects different abstraction levels: lightweight quantum estimators focus on minimal circuit construction, whereas transformer-based architectures demand extensive tokenization for parameter definitions, layer configurations, and operator chaining.

\begin{table*}[h]
\vspace{-2mm}
\caption{Statistical overview of CML-2-QML structure and paradigm distribution. E. denotes *extension*, while C. denotes *combination*.}
\vspace{-2mm}
\begin{adjustbox}{width=0.99\textwidth,center}
\begin{tabular}{l|c|cccc|cc|c}
\toprule
\multirow{2}{*}{\textbf{Seed Codebase}} & \multirow{2}{*}{\textbf{Ave. Length}} & \multicolumn{4}{c|}{\textbf{Reference}} & \multicolumn{2}{c|}{\textbf{Paradigm}} & \multirow{2}{*}{\textbf{Total}} \\
 &  & 1 & 2 & 3 & 4 & E. & C. & \\
\midrule
\midrule
Autoencoder & 3232 & 139 & 276 & 114 & 207 & 139 & 597 & 736 \\
ClassicalQuantumBinaryClassification & 3568 & 147 & 263 & 100 & 206 & 147 & 569 & 716 \\
Conv & 1490 & 160 & 261 & 109 & 232 & 160 & 602 & 762 \\
EstimatorQNN & 754 & 157 & 309 & 104 & 252 & 157 & 665 & 822 \\
FastBaseEstimator & 1926 & 160 & 266 & 113 & 226 & 160 & 605 & 765 \\
FCL & 1068 & 153 & 263 & 102 & 230 & 153 & 595 & 748 \\
FraudDetection & 2182 & 147 & 274 & 107 & 240 & 147 & 621 & 768 \\
GraphQNN & 4013 & 138 & 228 & 88 & 191 & 138 & 507 & 645 \\
QCNN & 2944 & 151 & 272 & 95 & 196 & 151 & 563 & 714 \\
QLSTM & 3756 & 135 & 277 & 91 & 192 & 135 & 560 & 695 \\
QTransformerTorch & 9628 & 136 & 229 & 48 & 150 & 136 & 427 & 563 \\
QuantumClassifierModel & 1096 & 159 & 286 & 102 & 214 & 159 & 602 & 761 \\
QuantumKernelMethod & 1722 & 149 & 291 & 103 & 226 & 149 & 620 & 751 \\
QuantumNAT & 1494 & 140 & 303 & 109 & 238 & 140 & 650 & 790 \\
QuantumRegression & 2162 & 152 & 298 & 113 & 196 & 152 & 607 & 759 \\
Quanvolution & 1558 & 143 & 292 & 104 & 241 & 143 & 637 & 780 \\
SamplerQNN & 777 & 157 & 266 & 91 & 236 & 157 & 593 & 750 \\
SelfAttention & 1258 & 176 & 314 & 107 & 255 & 176 & 676 & 852 \\
\midrule
\rowcolor[gray]{0.9} Total & 2697 & 2699 & 4968 & 1800 & 3828 & 2699 & 10696 & 13395 \\
\bottomrule
\end{tabular}
\end{adjustbox}
\label{tab:6}
\vspace{-2mm}
\end{table*}

The reference columns (1–4) collectively accumulate 13{,}395 entries, suggesting robust multi-level mapping for each seed code. Among them, reference~2 contributes the largest share (4{,}968), implying that intermediate-scale components are the most frequently reused or aligned during dataset construction. In the paradigm section, “E.” (extension) and “C.” (combination) categorize how each code sample expands or integrates existing patterns. The combination category dominates (10{,}696) over extension (2{,}699), indicating that most data transformations involve compositional reuse rather than independent structural growth. This imbalance suggests that quantum variants often build upon classical foundations through modular adaptation rather than entirely new formulations.

Collectively, these statistics confirm that the seed codebase achieves a balanced representation between simplicity and diversity, ensuring scalability for the CML-to-QML translation pipeline. The varying length distribution and cross-paradigm mapping make it suitable for instruction-tuning tasks that demand both fine-grained syntax alignment and conceptual transfer between classical and quantum machine learning domains.

\subsubsection{Case Studies}

\begin{figure}[H]
    \centering
    \includegraphics[width=0.98\linewidth]{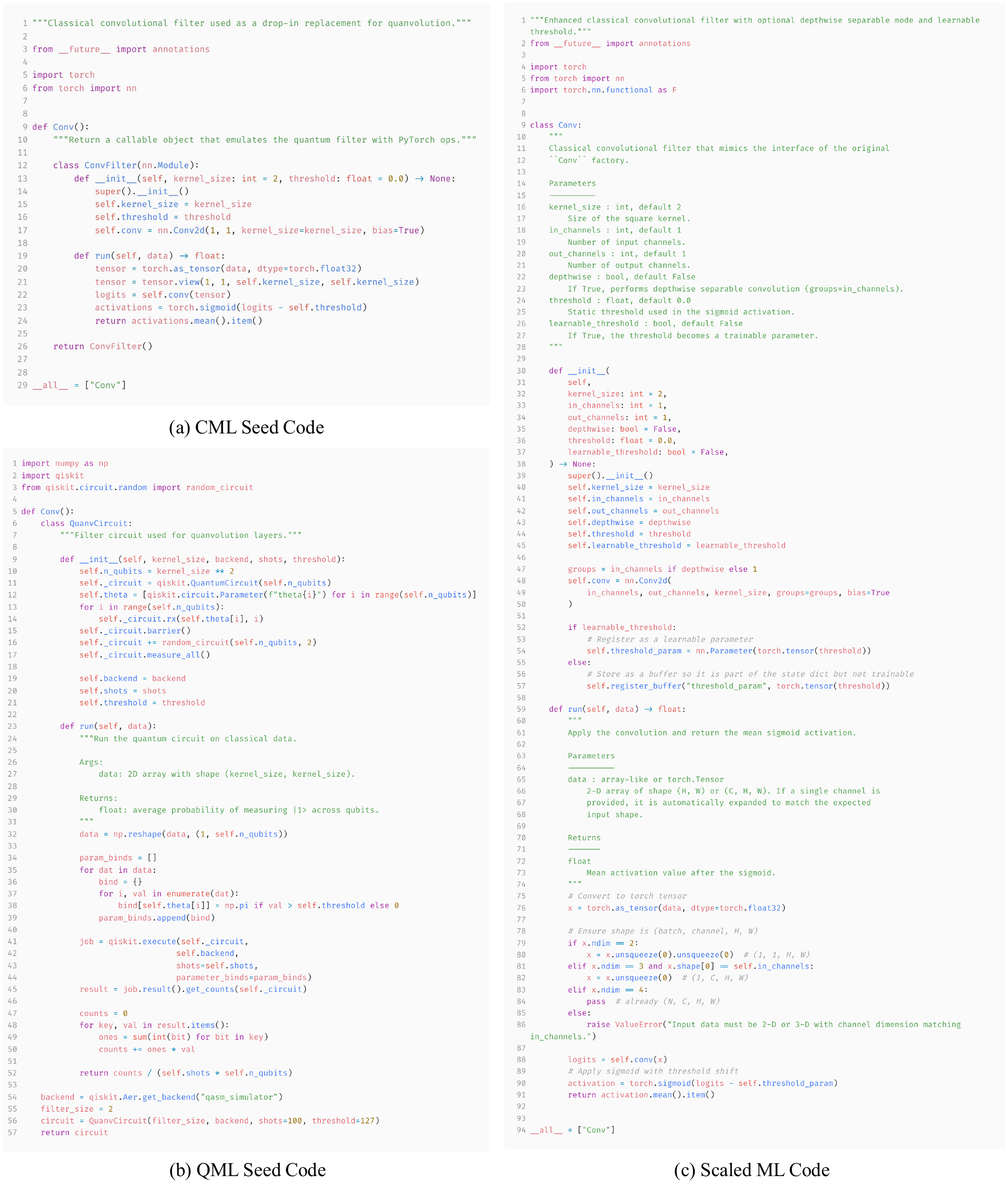}
    \vspace{-3mm}
    \caption{Comparison between seed code and its scaled ML version.} 
    \label{fig:4}
    \vspace{-4mm}
\end{figure}

As shown in Figure \ref{fig:4}, the convolutional ML and QML pair in the seed codebase have been refactored with concrete architectural and training upgrades. Two main differences are observed. First, the scaled code is generally longer than the original, incorporating more detailed comments and in-depth coding. Second, the implementation has shifted from Qiskit to Pennylane, expanding the diversity of supported quantum code libraries.

On the ML side, the module evolves from a single channel, fixed threshold convolutional factory into a parameterizable class that supports \textit{in\_channels}, \textit{out\_channels}, an optional depthwise mode, and a threshold that can be either learnable or stored as a registered buffer. It normalizes diverse input ranks to NCHW and preserves scalar activation through $\sigma(\text{logits}-\tau)$. These modifications expand representational capacity at modest parameter cost and align with efficient mobile style design. The learnable threshold improves calibration of the sigmoid operating point, and its registration as a parameter or buffer ensures checkpointing consistency. Enhanced input flexibility reduces brittle reshaping and supports microbatching.

\begin{figure}[H]
    \centering
    \includegraphics[width=1.0\linewidth]{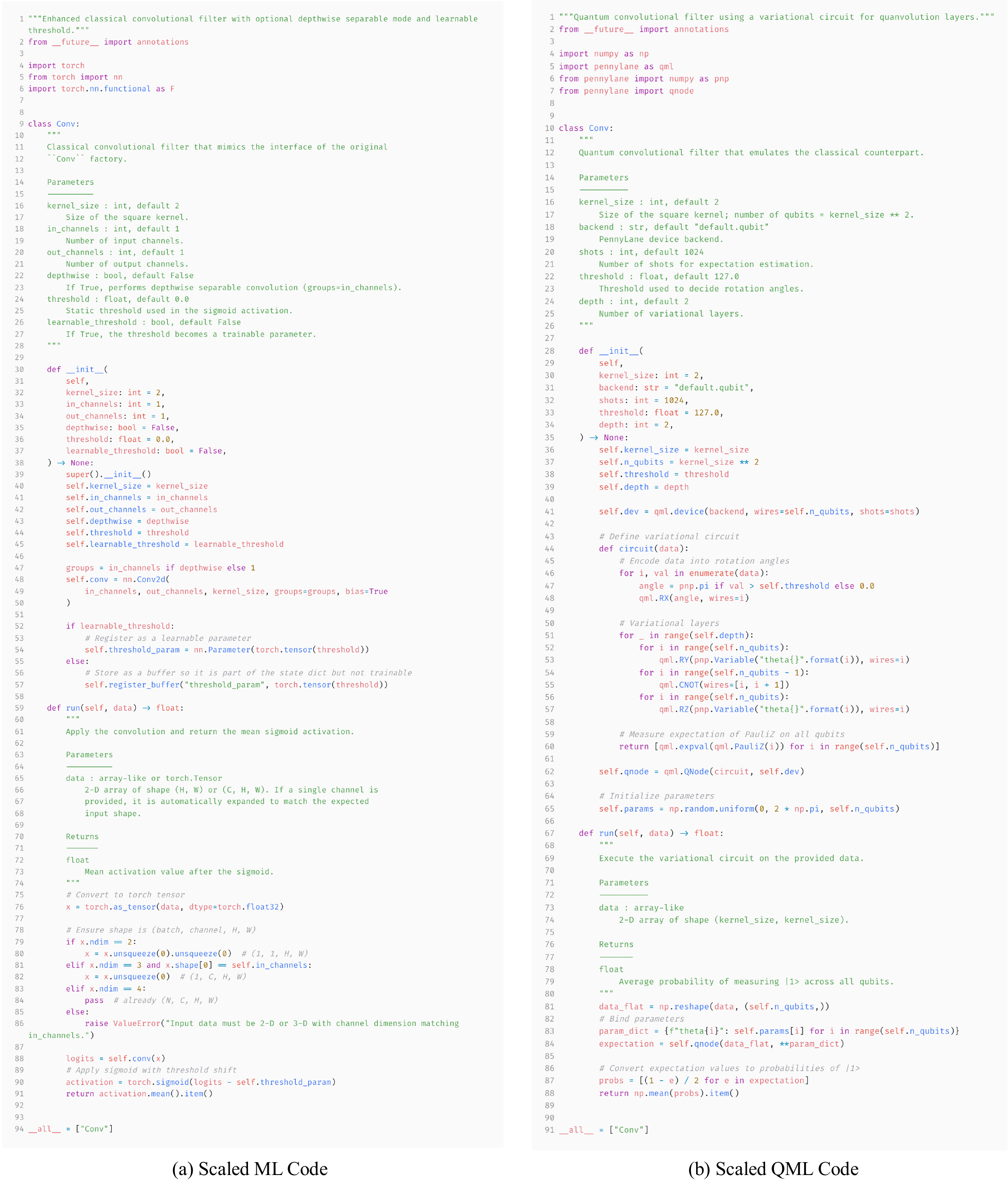}
    \vspace{-3mm}
    \caption{Case study of a paired generation of ML and QML code.} 
    \label{fig:5}
    \vspace{-4mm}
\end{figure}

As shown in Figure \ref{fig:5}, on the QML side, the static random circuit is replaced with a differentiable, parameter shifted variational ansatz that encodes data via $R_{X}$ gates, followed by layered $R_{Y}$, entangling CNOTs, and $R_{Z}$ gates with named trainable parameters, returning the mean probability of $\lvert 1\rangle$ across qubits. Both classes retain a unified \textit{run} interface that maps a 2D array to a scalar, with minor instantiation differences introducing limited API friction. This design enables gradient based learning with deterministic forward behavior, eliminating randomness and allowing end to end training across the hybrid boundary. The output ranges of both sides remain within $[0,1]$, stabilizing downstream modules expecting bounded scalar activations. Collectively, these changes transform the demonstrator pair into a trainable, research grade duo suitable for ablation studies on capacity, calibration, and hybrid co-adaptation.

Backward compatibility, however, is slightly weakened since the legacy factory pattern is not preserved. A thin \texttt{Conv()} wrapper that instantiates \texttt{Conv(...)} with legacy defaults could restore drop-in compatibility. Depthwise mode currently implements only the separable stage; adding an optional pointwise $1{\times}1$ projection would reintroduce cross channel mixing and improve performance on complex datasets. The learnable threshold remains global, \ie, enabling a channelwise option could enhance convergence under heterogeneous feature scales. In the scaled QML implementation, hard thresholding for $R_{X}$ selection is non-differentiable and discards magnitude information. Employing a smooth encoder such as $R_{X}(\alpha x)$ or normalized amplitude embedding would preserve data gradients. Further improvements include vectorizing parameter handling to reduce overhead and explicitly specifying device and shot count for reproducibility. A minimal training example with a loss function, optimizer, and parameter-shift loop would further lower integration costs and standardize evaluation across ML and QML baselines.

\subsection{Study of Q-Bridge Model}

Since our model is trained on the CML-2-QML dataset, which primarily consists of unverifiable code pairs, we conduct a detailed case study to qualitatively assess the translation behavior of our proposed Q-Bridge model against several strong baselines, including GPT-5-Thinking, GPT-OSS-20B, and Qwen3-1.7B. The case study uses CML code samples drawn from the held-out test set and applies an identical prompting template for all models to ensure fair comparison.

\begin{tcolorbox}[colback=lightgray!10, colframe=black,  title={Inference Prompt}]
\begin{Verbatim}[breaklines=true, breaksymbolleft={}, fontsize=\footnotesize]
You are an expert quantum machine learning researcher. Translate the provided classical machine learning (CML) description into its quantum machine learning (QML) counterpart.

CML Description: {cml_code}

QML Solution:
\end{Verbatim}
\end{tcolorbox}

As shown in Figure \ref{fig:6}, this is the test case used across all models. It translates a compact classical module \texttt{CML.py}, a flexible depth PyTorch classifier that takes a 2D input, applies Linear\,$\rightarrow$\,ReLU\,$\rightarrow$\,Dropout stacks, and outputs a 2-class softmax into an interpretable, executable quantum counterpart. The classical specification foregrounds the architectural motifs, \ie, two dimensional input, configurable hidden depth, dropout regularization, and probability outputs. The paired reference \texttt{QML.py} realizes these motifs as a 2-qubit PennyLane circuit, \ie, input features are encoded by $R_y$ rotations, entanglement by \texttt{CNOT}, followed by stacked trainable $R_y$ layers, and the circuit returns a probability distribution over computational basis states. We evaluate each output along four evaluation criterias: 

\begin{figure}[h]
    \centering
    \includegraphics[width=1.0\linewidth]{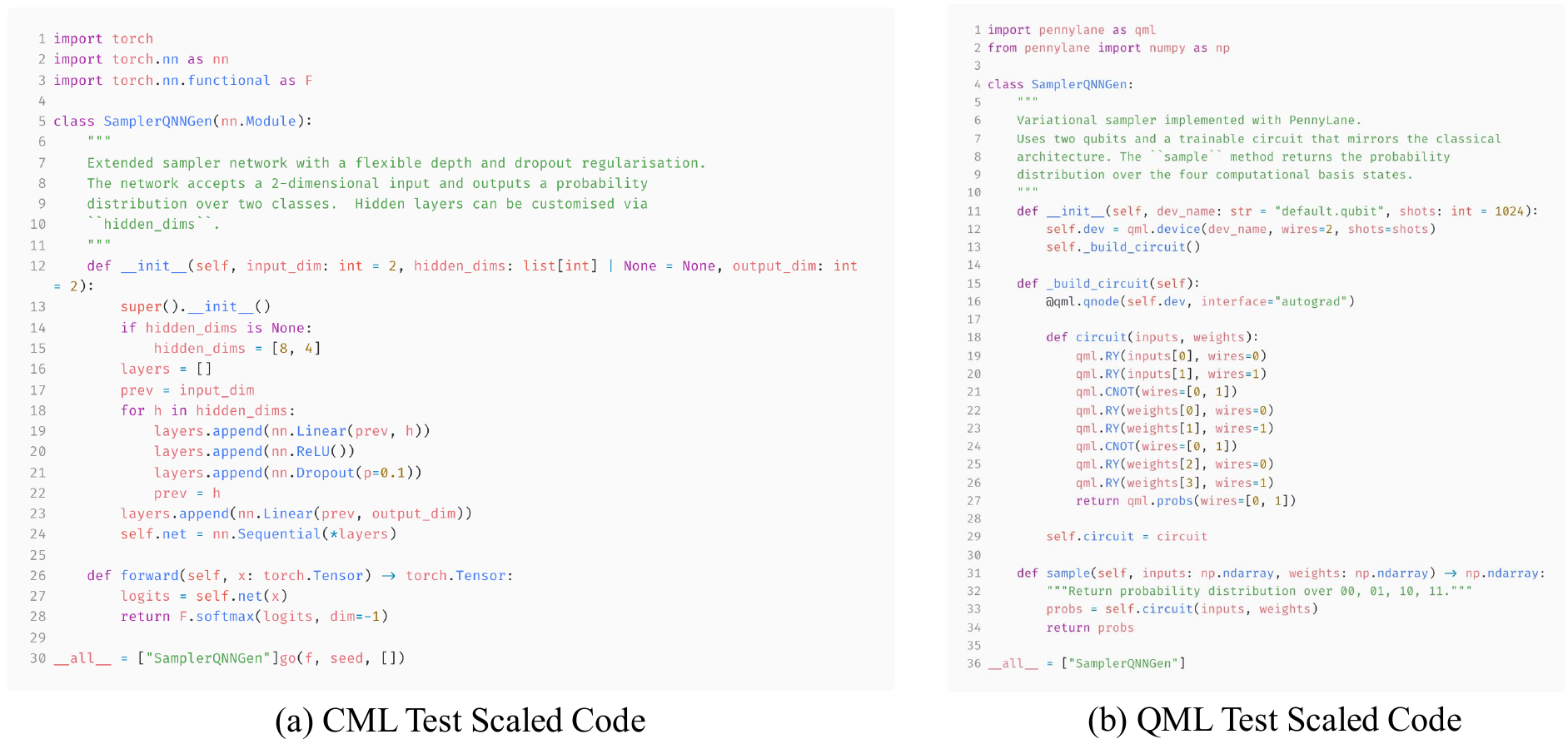}
    \vspace{-3mm}
    \caption{Test case of a paired generation of CML and QML code.} 
    \vspace{-2mm}
    \label{fig:6}
    \vspace{-2mm}
\end{figure}

(i) architectural fidelity: use two qubits for the 2D input; set the quantum circuit depth to mirror the classical model's hidden layer count, replace ReLU/Dropout with quantum equivalents (variational layers + randomized regularization), (ii) Quantum native design: use standard QML building blocks, \ie, parameterized single and two qubit gates, explicit feature maps/ans\"atz for data encoding and trainable layers, and (when appropriate) QNN wrappers (\eg, Estimator/Sampler) that expose a clear learning interface, (iii) executability and interpretability: the code should run in mainstream frameworks (\eg, Qiskit), separate data inputs from trainable weights, and structure the circuit into clearly named, reusable blocks so each layer’s role is transparent, (iv) alignment to the reference: preserve the reference architecture’s essentials, \ie, $R_y$ encoders for the two inputs, CX based entanglement between qubits, repeated variational layers that mirror the intended depth, and a probability distribution as the model output. 

To systematically examine the stochasticity and determinism of generation, we vary the temperature parameter, which controls the randomness of token sampling. For Q-Bridge, we evaluate three temperature levels: 0.1, 0.3, and 0.6. A temperature of 0.1 enforces near-deterministic decoding, emphasizing structural faithfulness and reproducibility of the generated QML code. 0.3 introduces moderate diversity, balancing syntactic consistency with exploratory token substitution—useful for discovering novel circuit or layer configurations. 0.6, the default used for all other baseline models, increases sampling entropy, promoting greater variety but potentially at the cost of semantic stability and logical coherence. This controlled temperature sweep allows us to analyze Q-Bridge’s adaptability under varying creativity–consistency trade-offs and to benchmark how its output quality and code correctness evolve relative to general-purpose large language models.

Baseline models underperform on fidelity, modularity, or reproducibility. GPT-5-Thinking adopts a sophisticated Qiskit ML stack (feature map + \texttt{RealAmplitudes} ansatz, \texttt{EstimatorQNN}, TorchConnector) and even emulates dropout via parameter masking. Yet it hard-codes a 2-class head, couples logits to an distinctive $\{+Z,-Z\}$ observable pair on a single qubit, and ties circuit depth to a brittle repetition heuristic. These choices impede extensibility and blur interpretability of the output head relative to the reference’s clear probability model. GPT-OSS-20B better mirrors the reference, \ie, it uses two wires, encodes inputs via $R_x$, stacks with \texttt{CNOT} entanglement, and returns probabilities, it injects probabilistically with a Pauli $X$ dropout. However, GPT-OSS-20B explicitly rejects non 2-D inputs and hard wires parameter shapes to a fixed qubit count constraining scalability and obscuring the mapping from the classical model’s hidden layer count to circuit depth, whereas Qwen3-1.7B provides only narrative text with no runnable circuit, failing all four criteria outright. 

\begin{figure}[H]
    \centering
    \includegraphics[width=1.0\linewidth]{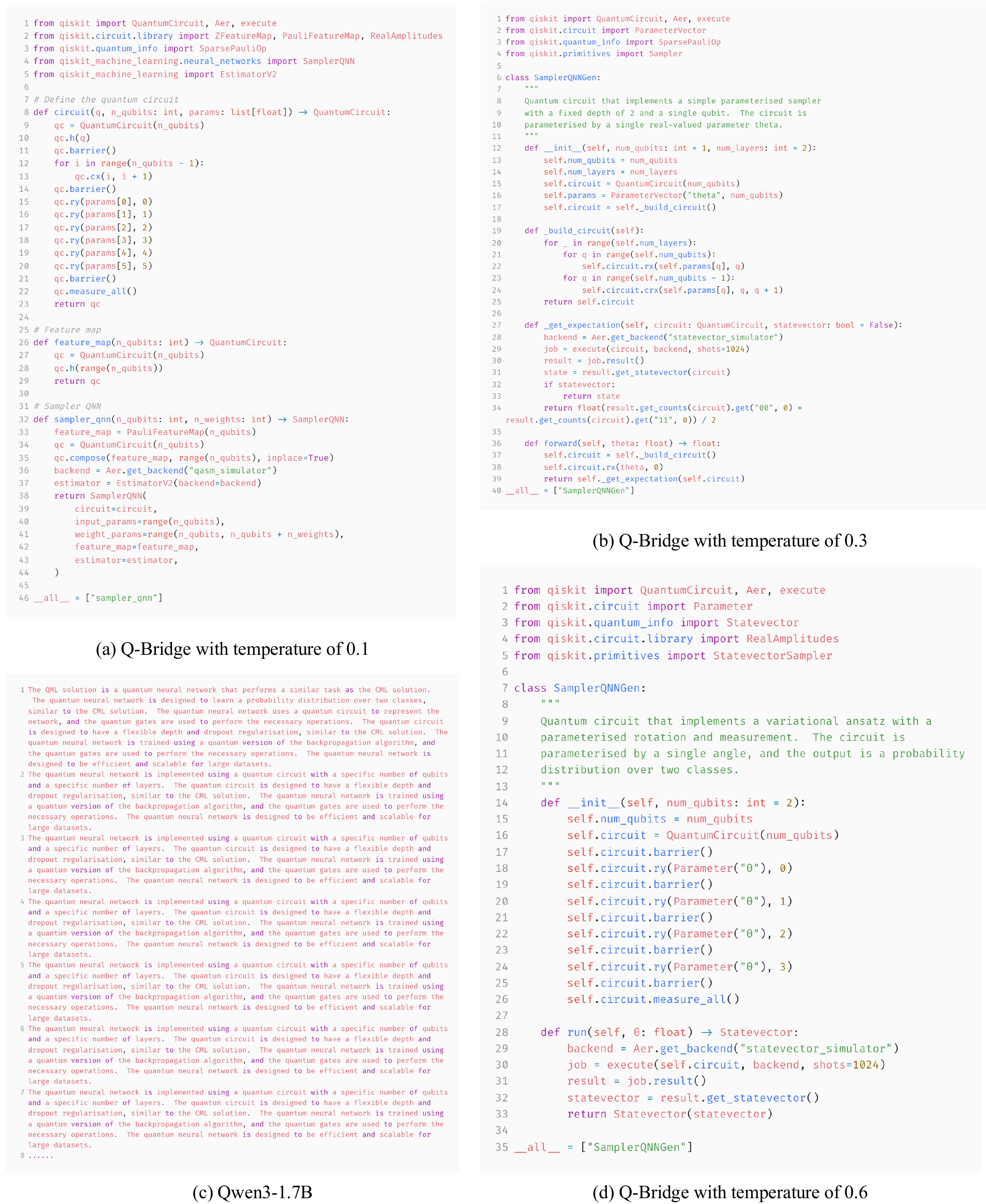}
    \vspace{-3mm}
    \caption{Generation results of Qwen3-1.7B and Q-Bridge under different temperatures.} 
    \label{fig:7}
    \vspace{-4mm}
\end{figure}

\begin{figure}[H]
    \centering
    \includegraphics[width=1.0\linewidth]{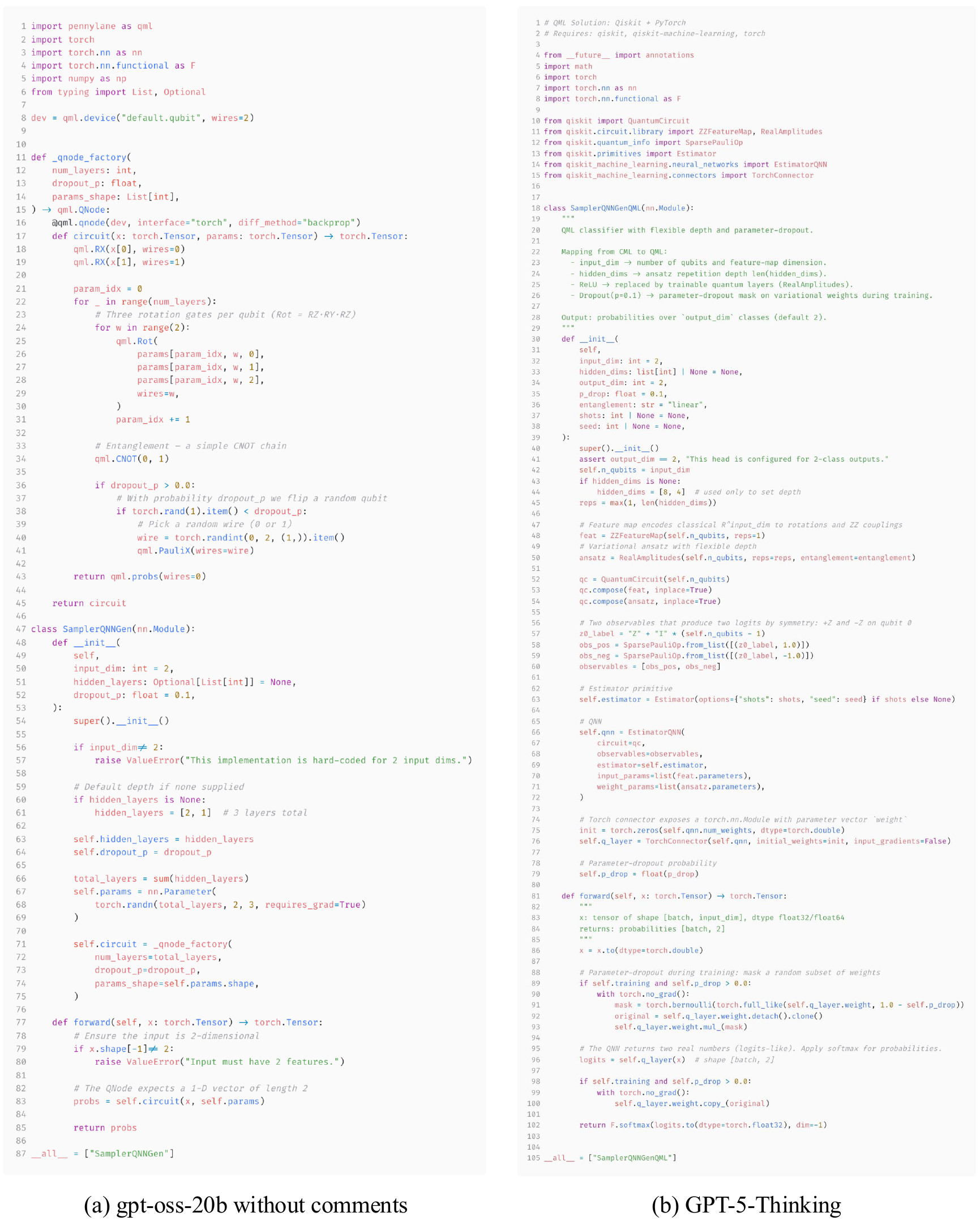}
    \vspace{-3mm}
    \caption{Generation results of baseline models.} 
    \label{fig:8}
    \vspace{-4mm}
\end{figure}

Collectively, the Q-Bridge variants more faithfully realize the task specification, employ cleaner quantum native structure, and yield reproducible, executable circuits, outperforming the GPT baselines in fidelity, modularity, and robustness. Q-Bridge-0-1 provides a minimal, first principles variational skeleton, \ie, explicit $R_y$ rotations with standard measurements, avoiding heavyweight QNN wrappers or connector stacks. This choice favors transparency and incremental extension. Although it uses a single shared angle across $R_y$ gates, the code cleanly exposes where to add feature encoding, entanglement, and extra layers so circuit depth can track the classical model’s hidden layer count. By contrast, GPT-5-Thinking couples feature maps, ansatz, and a custom observable head, complicating the logits pathway and making principled refactoring, and a clear mapping to depth significantly harder.

Q-Bridge-0-3 strengthens fidelity and modularity by introducing per qubit parameter vectors, an explicit layer builder, and controlled entanglement via a \texttt{CRX} chain. It cleanly separates number of qubits, number of layers, and per qubit parameters, yielding a reusable interface unlike GPT-5-Thinking's weight vector and GPT-OSS-20B's rigid 2-D only input check. Consequently, 0-3 scales naturally in qubits and depth and makes the mapping from the classical hidden layer count to variational depth explicit. Its forward path computes a concrete observable from executed circuits, enabling reproducible evaluation capabilities absent from Qwen3-1.7B's non-executable narrative. Q-Bridge-0-6 meets all criteria. It cleanly factors the pipeline into (a) feature encoding (a Pauli feature map that maps the two input features to two qubits), (b) variational layers with entanglement (Hadamards with a CX chain and per qubit $R_y$ rotations), and (c) probability readout via a standard \texttt{SamplerQNN} with modern \texttt{EstimatorV2}. The design mirrors the reference while using canonical Qiskit ML components, yielding an interpretable, end to end probability model without GPT-5’s brittle observable setup or GPT-OSS’s hard coded constraints. Crucially, 0-6 preserves the classical alignment, \ie, 2D inputs $\mapsto$ two qubits with explicit feature mapping, hidden layer depth $\mapsto$ layered rotations under CX entanglement, and softmax style outputs $\mapsto$ a probability distribution returned directly by the QNN. Across all four criteria, the Q-Bridge family outperforms the GPT baselines. The early, intentionally minimal 0-1 variant prioritizes editability and transparent quantum primitives over opaque plumbing, 0-3 adds principled layering and parameter modularity, scaling cleanly and making explicit the mapping from the classical hidden layer count to quantum circuit depth, and 0-6 delivers the complete, reproducible stack feature mapping, entanglement, and direct probability readout aligned with the reference via standard QNN tooling. By contrast, GPT-5-Thinking's rigid observable head and connector coupling, GPT-OSS-20B's input shape rigidity, and Qwen3-1.7B's non executable prose fall short on fidelity, modularity, or reproducibility. Consequently, with regards to Q-Bridge, especially version 0-6 is the superior, solution for CML$\rightarrow$QML translation in this benchmark.

\section{Conclusion}

We present Q-Bridge, the first large-scale framework for LLM-based code translation from classical to quantum machine learning. Q-Bridge contributes a verified-and-scaled dataset (CML-2-QML), a LoRA-tuned translation model, and a systematic pipeline enabling reproducible and interpretable QML generation. Its strengths include: (i) data-centric scalability through verifiable and unverifiable code expansion; (ii) efficient fine-tuning with low-rank adaptation for evolving quantum libraries; and (iii) flexible temperature control for balancing determinism and creativity during inference. Future directions include parameter-efficient fine-tuning for dynamic API adaptation, model distillation and quantization for lightweight deployment, and continual dataset growth for enhanced generalization. Together, these results demonstrate a practical path toward automated, explainable, and scalable quantum code generation. 
\bibliographystyle{ACM-Reference-Format}
\bibliography{main}

@inproceedings{zeng2026tokenseek,
  title={TokenSeek: Memory Efficient Fine Tuning via Instance-Aware Token Ditching},
  author={Zeng, Runjia and Wang, Qifan and Guan, Qiang and Tang, Ruixiang and Huang, Lifu and Wang, Zhenting and Zhang, Xueling and Han, Cheng and Liu, Dongfang},
  booktitle={ICLR},
  year={2026}
}

@inproceedings{zeng2025probabilistic,
  title={Probabilistic Token Alignment for Large Language Model Fusion},
  author={Zeng, Runjia and Liang, James Chenhao and Han, Cheng and Cao, Zhiwen and Liu, Jiahao and Quan, Xiaojun and Chen, Yingjie Victor and Huang, Lifu and Geng, Tong and Wang, Qifan and Liu, Dongfang},
  booktitle={NeurIPS},
  year={2025}
}

@inproceedings{zeng2024visual,
  title={Visual Fourier Prompt Tuning},
  author={Zeng, Runjia and Han, Cheng and Wang, Qifan and Wu, Chunshu and Geng, Tong and Huang, Lifu and Wu, Ying Nian and Liu, Dongfang},
  booktitle={NeurIPS},
  year={2024}
}

@inproceedings{zeng2025mept,
  title={MEPT: Mixture of Expert Prompt Tuning as a Manifold Mapper},
  author={Zeng, Runjia and Sun, Guangyan and Wang, Qifan and Geng, Tong and Dianat, Sohail and Han, Xiaotian and Rao, Raghuveer and Zhang, Xueling and Han, Cheng and Huang, Lifu and others},
  booktitle={EMNLP},
  year={2025}
}

@article{agarwal2025gpt,
  title={gpt-oss-120b \& gpt-oss-20b model card},
  author={Agarwal, Sandhini and Ahmad, Lama and Ai, Jason and Altman, Sam and Applebaum, Andy and Arbus, Edwin and Arora, Rahul K and Bai, Yu and Baker, Bowen and Bao, Haiming and others},
  journal={arXiv preprint arXiv:2508.10925},
  year={2025}
}

@article{haluptzok2022language,
  title={Language models can teach themselves to program better},
  author={Haluptzok, Patrick and Bowers, Matthew and Kalai, Adam Tauman},
  journal={arXiv preprint arXiv:2207.14502},
  year={2022}
}

@article{wang2023codet5+,
  title={Codet5+: Open code large language models for code understanding and generation},
  author={Wang, Yue and Le, Hung and Gotmare, Akhilesh Deepak and Bui, Nghi DQ and Li, Junnan and Hoi, Steven CH},
  journal={arXiv preprint arXiv:2305.07922},
  year={2023}
}

@article{chen2021evaluating,
  title={Evaluating large language models trained on code},
  author={Chen, Mark and Tworek, Jerry and Jun, Heewoo and Yuan, Qiming and Pinto, Henrique Ponde De Oliveira and Kaplan, Jared and Edwards, Harri and Burda, Yuri and Joseph, Nicholas and Brockman, Greg and others},
  journal={arXiv preprint arXiv:2107.03374},
  year={2021}
}

@article{nijkamp2022codegen,
  title={Codegen: An open large language model for code with multi-turn program synthesis},
  author={Nijkamp, Erik and Pang, Bo and Hayashi, Hiroaki and Tu, Lifu and Wang, Huan and Zhou, Yingbo and Savarese, Silvio and Xiong, Caiming},
  journal={arXiv preprint arXiv:2203.13474},
  year={2022}
}

@article{feng2020codebert,
  title={Codebert: A pre-trained model for programming and natural languages},
  author={Feng, Zhangyin and Guo, Daya and Tang, Duyu and Duan, Nan and Feng, Xiaocheng and Gong, Ming and Shou, Linjun and Qin, Bing and Liu, Ting and Jiang, Daxin and others},
  journal={arXiv preprint arXiv:2002.08155},
  year={2020}
}

@article{yu2024humaneval,
  title={Humaneval pro and mbpp pro: Evaluating large language models on self-invoking code generation},
  author={Yu, Zhaojian and Zhao, Yilun and Cohan, Arman and Zhang, Xiao-Ping},
  journal={arXiv preprint arXiv:2412.21199},
  year={2024}
}

@article{sirovs2024github,
  title={Github copilot: the perfect code compleeter?},
  author={Siro{\v{s}}, Ilja and Singel{\'e}e, Dave and Preneel, Bart},
  journal={arXiv preprint arXiv:2406.11326},
  year={2024}
}

@article{li2023starcoder,
  title={Starcoder: may the source be with you!},
  author={Li, Raymond and Allal, Loubna Ben and Zi, Yangtian and Muennighoff, Niklas and Kocetkov, Denis and Mou, Chenghao and Marone, Marc and Akiki, Christopher and Li, Jia and Chim, Jenny and others},
  journal={arXiv preprint arXiv:2305.06161},
  year={2023}
}

@inproceedings{khan2022automatic,
  title={Automatic code documentation generation using gpt-3},
  author={Khan, Junaed Younus and Uddin, Gias},
  booktitle={Proceedings of the 37th IEEE/ACM International Conference on Automated Software Engineering},
  pages={1--6},
  year={2022}
}

@article{wen2022code2tree,
  title={Code2tree: a method for automatically generating code comments},
  author={Wen, Wanzhi and Chu, Jiawei and Zhao, Tian and Zhang, Ruinian and Zhi, Bao and Shen, Chenqiang},
  journal={Scientific Programming},
  volume={2022},
  number={1},
  pages={6350686},
  year={2022},
  publisher={Wiley Online Library}
}

@inproceedings{makharev2025code,
  title={Code Summarization Beyond Function Level},
  author={Makharev, Vladimir and Ivanov, Vladimir},
  booktitle={2025 IEEE/ACM International Workshop on Large Language Models for Code (LLM4Code)},
  pages={153--160},
  year={2025},
  organization={IEEE}
}

@article{hong2025retrieval,
  title={Retrieval-Augmented Code Review Comment Generation},
  author={Hong, Hyunsun and Baik, Jongmoon},
  journal={arXiv preprint arXiv:2506.11591},
  year={2025}
}

@article{gong2024ast,
  title={Ast-t5: Structure-aware pretraining for code generation and understanding},
  author={Gong, Linyuan and Elhoushi, Mostafa and Cheung, Alvin},
  journal={arXiv preprint arXiv:2401.03003},
  year={2024}
}

@inproceedings{zhang2025function,
  title={Function-to-style guidance of llms for code translation},
  author={Zhang, Longhui and Wang, Bin and Wang, Jiahao and Zhao, Xiaofeng and Zhang, Min and Yang, Hao and Zhang, Meishan and Li, Yu and Li, Jing and Yu, Jun},
  booktitle={Forty-second International Conference on Machine Learning},
  year={2025}
}

@article{ibrahimzada2025alphatrans,
  title={AlphaTrans: A Neuro-Symbolic Compositional Approach for Repository-Level Code Translation and Validation},
  author={Ibrahimzada, Ali Reza and Ke, Kaiyao and Pawagi, Mrigank and Abid, Muhammad Salman and Pan, Rangeet and Sinha, Saurabh and Jabbarvand, Reyhaneh},
  journal={Proceedings of the ACM on Software Engineering},
  volume={2},
  number={FSE},
  pages={2454--2476},
  year={2025},
  publisher={ACM New York, NY, USA}
}

@article{li2024few,
  title={Few-shot code translation via task-adapted prompt learning},
  author={Li, Xuan and Yuan, Shuai and Gu, Xiaodong and Chen, Yuting and Shen, Beijun},
  journal={Journal of Systems and Software},
  volume={212},
  pages={112002},
  year={2024},
  publisher={Elsevier}
}

@article{joos2025codecureagent,
  title={CodeCureAgent: Automatic Classification and Repair of Static Analysis Warnings},
  author={Joos, Pascal and Bouzenia, Islem and Pradel, Michael},
  journal={arXiv preprint arXiv:2509.11787},
  year={2025}
}

@article{jiang2024survey,
  title={A survey on large language models for code generation},
  author={Jiang, Juyong and Wang, Fan and Shen, Jiasi and Kim, Sungju and Kim, Sunghun},
  journal={arXiv preprint arXiv:2406.00515},
  year={2024}
}

@article{cerezo2021variational,
  title={Variational quantum algorithms},
  author={Cerezo, Marco and Arrasmith, Andrew and Babbush, Ryan and Benjamin, Simon C and Endo, Suguru and Fujii, Keisuke and McClean, Jarrod R and Mitarai, Kosuke and Yuan, Xiao and Cincio, Lukasz and others},
  journal={Nature Reviews Physics},
  volume={3},
  number={9},
  pages={625--644},
  year={2021},
  publisher={Nature Publishing Group UK London}
}

@article{peruzzo2014variational,
  title={A variational eigenvalue solver on a photonic quantum processor},
  author={Peruzzo, A. and McClean, J. and Shadbolt, P. and Yung, M.H. and Zhou, X.Q. and Love, P.J. and Aspuru-Guzik, A. and O'Brien, J.L.},
  journal={Nature Communications},
  volume={5},
  pages={4213},
  year={2014},
  publisher={Nature Publishing Group}
}

@article{farhi2022quantum,
  title={The quantum approximate optimization algorithm and the Sherrington-Kirkpatrick model at infinite size},
  author={Farhi, Edward and Goldstone, Jeffrey and Gutmann, Sam and Zhou, Leo},
  journal={Quantum},
  volume={6},
  pages={759},
  year={2022},
  publisher={Verein zur F{\"o}rderung des Open Access Publizierens in den Quantenwissenschaften}
}

@article{mcclean2018barren,
  title={Barren Plateaus in Quantum Neural Network Training Landscapes},
  author={McClean, Jarrod R and Boixo, Sergio and Smelyanskiy, Vadim N and Babbush, Ryan and Neven, Hartmut},
  journal={Nature Communications},
  volume={9},
  number={1},
  pages={4812},
  year={2018},
  publisher={Nature Publishing Group}
}

@article{havlivcek2019supervised,
  title={Supervised learning with quantum-enhanced feature spaces},
  author={Havl{\'\i}{\v{c}}ek, Vojt{\v{e}}ch and C{\'o}rcoles, Antonio D and Temme, Kristan and Harrow, Aram W and Kandala, Abhinav and Chow, Jerry M and Gambetta, Jay M},
  journal={Nature},
  volume={567},
  number={7747},
  pages={209--212},
  year={2019},
  publisher={Nature Publishing Group}
}

@article{biamonte2017quantum,
  title={Quantum machine learning},
  author={Biamonte, Jacob and Wittek, Peter and Pancotti, Nicola and Rebentrost, Patrick and Wiebe, Nathan and Lloyd, Seth},
  journal={Nature},
  volume={549},
  number={7671},
  pages={195--202},
  year={2017},
  publisher={Nature Publishing Group UK London}
}

@article{scikit-learn,
  title={Scikit-learn: Machine Learning in {P}ython},
  author={Pedregosa, F. and Varoquaux, G. and Gramfort, A. and Michel, V.
          and Thirion, B. and Grisel, O. and Blondel, M. and Prettenhofer, P.
          and Weiss, R. and Dubourg, V. and Vanderplas, J. and Passos, A. and
          Cournapeau, D. and Brucher, M. and Perrot, M. and Duchesnay, E.},
  journal={Journal of Machine Learning Research},
  volume={12},
  pages={2825--2830},
  year={2011}
}

@article{sim2019expressibility,
  title={Expressibility and entangling capability of parameterized quantum circuits for hybrid quantum-classical algorithms},
  author={Sim, Sukin and Johnson, Peter D and Aspuru-Guzik, Al{\'a}n},
  journal={Advanced Quantum Technologies},
  volume={2},
  number={12},
  pages={1900070},
  year={2019},
  publisher={Wiley Online Library}
}

@misc{torchquantum,
  author = {Hanrui Wang},
  title = {Torch Quantum},
  year = {2021},
  publisher = {GitHub},
  journal = {GitHub repository},
  howpublished = {\url{https://github.com/mit-han-lab/torchquantum},
%   commit = {4f57d6a0e4c030202a07a60bc1bb1ed1544bf679}
}}

@misc{qiskit_human_eval,
Author = {Sanjay Vishwakarma and Francis Harkins and Siddharth Golecha and Vishal Sharathchandra Bajpe and Nicolas Dupuis and Luca Buratti and David Kremer and Ismael Faro and Ruchir Puri and Juan Cruz-Benito},
Title = {Qiskit HumanEval: An Evaluation Benchmark For Quantum Code Generative Models},
Year = {2024},
Eprint = {arXiv:2406.14712},
}

@inproceedings{dupuis2024qiskit,
  title={Qiskit code assistant: Training llms for generating quantum computing code},
  author={Dupuis, Nicolas and Buratti, Luca and Vishwakarma, Sanjay and Forrat, Aitana Viudes and Kremer, David and Faro, Ismael and Puri, Ruchir and Cruz-Benito, Juan},
  booktitle={2024 IEEE LLM Aided Design Workshop (LAD)},
  pages={1--4},
  year={2024},
  organization={IEEE}
}

@inproceedings{shor1994algorithms,
  title={Algorithms for quantum computation: discrete logarithms and factoring},
  author={Shor, Peter W},
  booktitle={Proceedings 35th annual symposium on foundations of computer science},
  pages={124--134},
  year={1994},
  organization={Ieee}
}

@inproceedings{grover1996fast,
  title={A fast quantum mechanical algorithm for database search},
  author={Grover, Lov K},
  booktitle={Proceedings of the twenty-eighth annual ACM symposium on Theory of computing},
  pages={212--219},
  year={1996}
}

@article{preskill2018quantum,
  title={Quantum computing in the NISQ era and beyond},
  author={Preskill, John},
  journal={Quantum},
  volume={2},
  pages={79},
  year={2018},
  publisher={Verein zur F{\"o}rderung des Open Access Publizierens in den Quantenwissenschaften}
}

@article{arute2019quantum,
  title={Quantum supremacy using a programmable superconducting processor},
  author={Arute, Frank and Arya, Kunal and Babbush, Ryan and Bacon, Dave and Bardin, Joseph C and Barends, Rami and Biswas, Rupak and Boixo, Sergio and Brandao, Fernando GSL and Buell, David A and others},
  journal={Nature},
  volume={574},
  number={7779},
  pages={505--510},
  year={2019},
  publisher={Nature Publishing Group UK London}
}

@article{caro2023out,
  title={Out-of-distribution generalization for learning quantum dynamics},
  author={Caro, Matthias C and Huang, Hsin-Yuan and Ezzell, Nicholas and Gibbs, Joe and Sornborger, Andrew T and Cincio, Lukasz and Coles, Patrick J and Holmes, Zo{\"e}},
  journal={Nature Communications},
  volume={14},
  number={1},
  pages={3751},
  year={2023},
  publisher={Nature Publishing Group UK London}
}

@article{caro2021encoding,
  title={Encoding-dependent generalization bounds for parametrized quantum circuits},
  author={Caro, Matthias C and Gil-Fuster, Elies and Meyer, Johannes Jakob and Eisert, Jens and Sweke, Ryan},
  journal={Quantum},
  volume={5},
  pages={582},
  year={2021},
  publisher={Verein zur F{\"o}rderung des Open Access Publizierens in den Quantenwissenschaften}
}

@inproceedings{aloisiod2024exploring,
  title={Exploring LLM-driven explanations for quantum algorithms},
  author={d'Aloisio, Giordano and Fortz, Sophie and Hanna, Carol and Fortunato, Daniel and Bensoussan, Avner and Mendiluze Usandizaga, E{\~n}aut and Sarro, Federica},
  booktitle={Proceedings of the 18th ACM/IEEE International Symposium on Empirical Software Engineering and Measurement},
  pages={475--481},
  year={2024}
}

@inproceedings{easttom2024utilizing,
  title={Utilizing ChatGPT to Improve Quantum Algorithms},
  author={Easttom, Chuck},
  booktitle={2024 IEEE 14th Annual Computing and Communication Workshop and Conference (CCWC)},
  pages={0744--0749},
  year={2024},
  organization={IEEE}
}

@misc{guo2024repairing,
  title={On Repairing Quantum Programs Using ChatGPT. In 2024 IEEE/ACM 5th International Workshop on Quantum Software Engineering (Q-SE)},
  author={Guo, Xiaoyu and Zhao, Jianjun and Zhao, Pengzhan},
  year={2024},
  publisher={IEEE}
}

@article{ibmfutureroadmap,
  title={IBM’s Roadmap For Scaling Quantum Technology},
  author={IBM},
  journal={https://www.ibm.com/roadmaps/quantum/},
  year={2025},
  note = {Accessed: 2025-10-17}
}

@inproceedings{senapati2024pqml,
  author={Senapati, Priyabrata and Chen, Samuel Yen-Chi and Fang, Bo and Athawale, Tushar M. and Li, Ang and Jiang, Weiwen and Lu, Cheng Chang and Guan, Qiang},
  booktitle={2024 IEEE International Conference on Quantum Computing and Engineering (QCE)}, 
  title={PQML: Enabling the Predictive Reproducibility on NISQ Machines for Quantum ML Applications}, 
  year={2024},
  volume={01},
  number={},
  pages={1413-1424},
  doi={10.1109/QCE60285.2024.00168}}

@inproceedings{jern2025agent,
  title={Agent-Q: Fine-Tuning Large Language Models for Quantum Circuit Generation and Optimization},
  author={Jern, Linus and Uotila, Valter and Yu, Cong and Zhao, Bo},
  booktitle={2025 IEEE International Conference on Quantum Computing and Engineering (QCE). IEEE},
  year={2025}
}

@inproceedings{xiong2025reinvent,
  title={Reinvent the Operation not the Architecture: Quantum-inspired High-order Product for Compatible and Improved LLMs Training},
  author={Xiong, Hao and Yang, Yebin and Wu, Huaijin and Zhong, Xiaoqiu and Tang, Yehui and Xia, Zhuo and Wang, Xiaoxing and Yan, Junchi},
  booktitle={Proceedings of the 31st ACM SIGKDD Conference on Knowledge Discovery and Data Mining V. 2},
  pages={3356--3365},
  year={2025}
}

@article{zhou2025application,
  title={Application of large language models to quantum state simulation},
  author={Zhou, Shuangxiang and Chen, Ronghang and An, Zheng and Zhang, Chao and Hou, Shi-Yao},
  journal={Science China Physics, Mechanics \& Astronomy},
  volume={68},
  number={4},
  pages={240313},
  year={2025},
  publisher={Springer}
}

@article{melko2024language,
  title={Language models for quantum simulation},
  author={Melko, Roger G and Carrasquilla, Juan},
  journal={Nature Computational Science},
  volume={4},
  number={1},
  pages={11--18},
  year={2024},
  publisher={Nature Publishing Group US New York}
}

@article{schuld2021supervised,
  title={Quantum machine learning in feature Hilbert spaces},
  author={Schuld, Maria and Killoran, Nathan},
  journal={Physical review letters},
  volume={122},
  number={4},
  pages={040504},
  year={2019},
  publisher={APS}
}

@article{cong2019quantum,
  title={Quantum convolutional neural networks},
  author={Cong, Iris and Choi, Soonwon and Lukin, Mikhail D},
  journal={Nature Physics},
  volume={15},
  number={12},
  pages={1273--1278},
  year={2019},
  publisher={Nature Publishing Group UK London}
}

@article{perez2020data,
  title={Data re-uploading for a universal quantum classifier},
  author={P{\'e}rez-Salinas, Adri{\'a}n and Cervera-Lierta, Alba and Gil-Fuster, Elies and Latorre, Jos{\'e} I},
  journal={Quantum},
  volume={4},
  pages={226},
  year={2020},
  publisher={Verein zur F{\"o}rderung des Open Access Publizierens in den Quantenwissenschaften}
}

@article{huang2021power,
  title={Power of data in quantum machine learning},
  author={Huang, Hsin-Yuan and Broughton, Michael and Mohseni, Masoud and Babbush, Ryan and Boixo, Sergio and Neven, Hartmut and McClean, Jarrod R},
  journal={Nature communications},
  volume={12},
  number={1},
  pages={2631},
  year={2021},
  publisher={Nature Publishing Group UK London}
}

@article{killoran2019continuous,
  title={Continuous-variable quantum neural networks},
  author={Killoran, Nathan and Bromley, Thomas R and Arrazola, Juan Miguel and Schuld, Maria and Quesada, Nicol{\'a}s and Lloyd, Seth},
  journal={Physical Review Research},
  volume={1},
  number={3},
  pages={033063},
  year={2019},
  publisher={APS}
}

@article{henderson2020quanvolutional,
  title={Quanvolutional neural networks: powering image recognition with quantum circuits},
  author={Henderson, Maxwell and Shakya, Samriddhi and Pradhan, Shashindra and Cook, Tristan},
  journal={Quantum Machine Intelligence},
  volume={2},
  number={1},
  pages={2},
  year={2020},
  publisher={Springer}
}

@article{beer2020training,
  title={Training deep quantum neural networks},
  author={Beer, Kerstin and Bondarenko, Dmytro and Farrelly, Terry and Osborne, Tobias J and Salzmann, Robert and Scheiermann, Daniel and Wolf, Ramona},
  journal={Nature communications},
  volume={11},
  number={1},
  pages={808},
  year={2020},
  publisher={Nature Publishing Group UK London}
}

@article{bondarenko2020quantum,
  title={Quantum autoencoders to denoise quantum data},
  author={Bondarenko, Dmytro and Feldmann, Polina},
  journal={Physical review letters},
  volume={124},
  number={13},
  pages={130502},
  year={2020},
  publisher={APS}
}

@inproceedings{wang2022torchquantum,
  title={Torchquantum case study for robust quantum circuits},
  author={Wang, Hanrui and Liang, Zhiding and Gu, Jiaqi and Li, Zirui and Ding, Yongshan and Jiang, Weiwen and Shi, Yiyu and Pan, David Z and Chong, Frederic T and Han, Song},
  booktitle={Proceedings of the 41st IEEE/ACM International Conference on Computer-Aided Design},
  pages={1--9},
  year={2022}
}

@inproceedings{quantumNAT,
author = {Wang, Hanrui and Gu, Jiaqi and Ding, Yongshan and Li, Zirui and Chong, Frederic T. and Pan, David Z. and Han, Song},
title = {QuantumNAT: quantum noise-aware training with noise injection, quantization and normalization},
year = {2022},
isbn = {9781450391429},
publisher = {Association for Computing Machinery},
address = {New York, NY, USA},
url = {https://doi.org/10.1145/3489517.3530400},
doi = {10.1145/3489517.3530400},
booktitle = {Proceedings of the 59th ACM/IEEE Design Automation Conference},
pages = {1–6},
numpages = {6},
location = {San Francisco, California},
series = {DAC '22}
}

@article{vatan2004optimal,
  title={Optimal quantum circuits for general two-qubit gates},
  author={Vatan, Farrokh and Williams, Colin},
  journal={Physical Review A—Atomic, Molecular, and Optical Physics},
  volume={69},
  number={3},
  pages={032315},
  year={2004},
  publisher={APS}
}

@INPROCEEDINGS{10821318,
  author={Meyer, Nico and Ufrecht, Christian and Periyasamy, Maniraman and Plinge, Axel and Mutschler, Christopher and Scherer, Daniel D. and Maier, Andreas},
  booktitle={2024 IEEE International Conference on Quantum Computing and Engineering (QCE)}, 
  title={Qiskit-Torch-Module: Fast Prototyping of Quantum Neural Networks}, 
  year={2024},
  volume={01},
  number={},
  pages={817-823},
  doi={10.1109/QCE60285.2024.00101}}

@misc{qiskit_ml_tutorial_qnn,
  title        = {Quantum Neural Networks (Tutorial 01)},
  author       = {{Qiskit Machine Learning Development Team}},
  year         = {2025},
  howpublished = {\url{https://qiskit-community.github.io/qiskit-machine-learning/tutorials/01_neural_networks.html}},
  note         = {Version 0.8.4. Jupyter notebook: \url{https://github.com/qiskit-community/qiskit-machine-learning/blob/stable/0.8/docs/tutorials/01_neural_networks.ipynb}},
  urldate      = {2025-10-26}
}






\end{document}